\documentclass[twocolumn,tighten]{aastex63}

\received{April 27, 2021}
\revised{May 27, 2021}
\accepted{June 27, 2021}
\submitjournal{ApJ}

\shorttitle{Deficient H$\alpha$ flux in the extended disks of spiral galaxies}
\shortauthors{Byun et al.}

\begin{document}

\title{KMTNet Nearby Galaxy Survey III. Deficient H$\alpha$ flux in the Extended Disks of Spiral Galaxies}

\correspondingauthor{Minjin Kim}
\email{mkim@knu.ac.kr}

\author[0000-0002-7762-7712]{Woowon Byun}
\affiliation{Korea Astronomy and Space Science Institute, Daejeon 34055, Republic of Korea}
\affiliation{University of Science and Technology, Korea, Daejeon 34113, Republic of Korea}

\author[0000-0002-3211-9431]{Yun-Kyeong Sheen}
\affiliation{Korea Astronomy and Space Science Institute, Daejeon 34055, Republic of Korea}

\author[0000-0001-9561-8134]{Kwang-Il Seon}
\affiliation{Korea Astronomy and Space Science Institute, Daejeon 34055, Republic of Korea}
\affiliation{University of Science and Technology, Korea, Daejeon 34113, Republic of Korea}

\author[0000-0001-6947-5846]{Luis C. Ho}
\affiliation{Kavli Institute for Astronomy and Astrophysics, Peking University, Beijing 100871, People's Republic of China}
\affiliation{Department of Astronomy, School of Physics, Peking University, Beijing 100871, People's Republic of China}

\author[0000-0003-3451-0925]{Joon Hyeop Lee}
\affiliation{Korea Astronomy and Space Science Institute, Daejeon 34055, Republic of Korea}
\affiliation{University of Science and Technology, Korea, Daejeon 34113, Republic of Korea}

\author[0000-0002-0145-9556]{Hyunjin Jeong}
\affiliation{Korea Astronomy and Space Science Institute, Daejeon 34055, Republic of Korea}

\author[0000-0001-9670-1546]{Sang Chul Kim}
\affiliation{Korea Astronomy and Space Science Institute, Daejeon 34055, Republic of Korea}
\affiliation{University of Science and Technology, Korea, Daejeon 34113, Republic of Korea}

\author[0000-0002-6982-7722]{Byeong-Gon Park}
\affiliation{Korea Astronomy and Space Science Institute, Daejeon 34055, Republic of Korea}
\affiliation{University of Science and Technology, Korea, Daejeon 34113, Republic of Korea}

\author{Yongseok Lee}
\affiliation{Korea Astronomy and Space Science Institute, Daejeon 34055, Republic of Korea}
\affiliation{School of Space Research, Kyung Hee University, Yongin, Kyeonggi 17104, Republic of Korea}

\author{Sang-Mok Cha}
\affiliation{Korea Astronomy and Space Science Institute, Daejeon 34055, Republic of Korea}
\affiliation{School of Space Research, Kyung Hee University, Yongin, Kyeonggi 17104, Republic of Korea}

\author[0000-0002-9434-5936]{Jongwan Ko}
\affiliation{Korea Astronomy and Space Science Institute, Daejeon 34055, Republic of Korea}
\affiliation{University of Science and Technology, Korea, Daejeon 34113, Republic of Korea}

\author[0000-0002-3560-0781]{Minjin Kim}
\affiliation{Department of Astronomy and Atmospheric Sciences, Kyungpook National University, Daegu 41566, Republic of Korea}




\begin{abstract}

We perform a deep wide-field imaging survey of nearby galaxies using 
H$\alpha$ and broadband filters to investigate the characteristics of star 
formation in galaxies. Motivated by the finding that star formation rates 
(SFRs) derived from H$\alpha$ fluxes in dwarf galaxies are systematically lower 
than those inferred from far-ultraviolet (FUV) fluxes, we attempt 
to determine whether the same trend exists in the extended disks of two 
star-forming galaxies. We perform spatially resolved photometry using 
grid-shaped apertures to measure the FUV and H$\alpha$ 
fluxes of star-forming regions. We also perform spectral energy distribution 
(SED) fittings using 11 photometric data (FUV-to-MIR) including data 
from the literature to estimate the local properties such as internal attenuation 
of individual star-forming clumps. Comparing SFR$_\mathrm{FUV}$ and 
SFR$_\mathrm{H\alpha}$, which are converted from the H$\alpha$ and FUV 
fluxes corrected for the local properties, we find that 
SFR$_\mathrm{H\alpha}$/SFR$_\mathrm{FUV}$ tends to decrease as the 
SFR decreases. We evaluate possible causes of this discrepancy between 
the two SFRs by restricting parameters in the SED fitting and conclude that 
deficient H$\alpha$ fluxes in the extended disks of galaxies are tightly correlated 
with recent starbursts. The strong and short starburst which is 
being rapidly suppressed over the last 10 Myr seems to induce a significant 
discrepancy between the SFR$_\mathrm{H\alpha}$ and SFR$_\mathrm{FUV}$. 
In addition, the recent bursts in the extended disks of galaxies appear to have 
occurred azimuth-symmetrically, implying that these were likely triggered by gas 
accretion or internal processes rather than external perturbation. 

\end{abstract}



\section{Introduction}\label{sec:intro}

H$\alpha$ emission line and far-ultraviolet (FUV) continuum fluxes 
have been widely used to estimate the star formation rate (SFR) of galaxies 
\citep[see][]{1998ARA&A..36..189K,2012ARA&A..50..531K}. H$\alpha$ 
emissions originate from ionized gas in the vicinity of the most massive stars 
such as O- and early B-stars, which have lifetimes of $\sim$10$^6$ yr. 
On the other hand, the FUV continuum is contributed even by less 
massive stars, such as late B-stars with much longer lifetimes of 
$\sim$10$^8$ yr. 

At a given stellar initial mass function (IMF), we can estimate the 
SFR$_\mathrm{FUV}$ and SFR$_\mathrm{H\alpha}$ using the FUV and H$\alpha$ fluxes, 
respectively. In principle, the resulting SFRs should be consistent with 
each other based on standard conversion recipes, which assume a constant ratio 
between FUV and H$\alpha$ fluxes \citep[e.g.,][]{1998ARA&A..36..189K}. 
However, it has been reported that low-mass dwarf galaxies 
tend to exhibit SFR$_\mathrm{H\alpha}$ lower than SFR$_\mathrm{FUV}$ 
\citep[e.g.,][]{2000MNRAS.312..442S,2001ApJ...548..681B,2009ApJ...706..599L, 
2016ApJ...817..177L}. In particular, \cite{2009ApJ...706..599L} conducted a 
systematic study using unbiased samples including dwarf galaxies, 
and concluded that the observed H$\alpha$-to-FUV flux ratio 
systematically decreases with decreasing luminosity. 

While the physical origin of this finding is unclear, 
previous studies have proposed several scenarios to explain it. 
For example, dwarf galaxies which have recently experienced a rapid 
suppression of bursty star formation can show a deficient H$\alpha$ flux due to the 
lack of very young stars, which generate ionizing photons \citep[e.g.,][]{2004MNRAS.350...21S,
2004A&A...421..887I,2012ApJ...744...44W,2019ApJ...881...71E}. It can also 
be explained by the leakage of ionizing photons from star-forming regions 
\citep[e.g.,][]{1997MNRAS.291..827O,2011MNRAS.411..235E,2012MNRAS.423.2933R}, 
suggesting that a supplementary detection of diffuse H$\alpha$ fluxes can mitigate 
the discrepancy \citep[e.g.,][]{2016ApJ...817..177L,2018A&A...611A..95W}. In addition, 
it has been suggested a steeper and/or truncated stellar IMF
 \citep[e.g.,][]{2009ApJ...695..765M,2009MNRAS.395..394P}, or a stochastic 
sampling effect may be responsible \citep[e.g.,][]{2011ApJ...741L..26F,2014MNRAS.444.3275D}. 

These studies on H$\alpha$ and FUV have mainly been conducted by 
using the entire light of individual galaxies \citep[e.g.,][]{1987A&A...185...33B,
2000MNRAS.312..442S,2001ApJ...548..681B,2004A&A...421..887I,
2007ApJS..173..267S}. We understand that the well-known global 
scaling relations, such as the mass-metallicity relation, can often be reproduced by 
spatially resolved properties even in a single galaxy \citep[e.g.,][]{2012ApJ...756L..31R,2013A&A...554A..58S,
2016MNRAS.463.2513B,2016ApJ...821L..26C,2017ApJ...851L..24H,
2020MNRAS.493.4107E,2020MNRAS.499.4838M,2020MNRAS.496.4606M,
2020ApJ...903...52S}. Similarly, the decrease in the H$\alpha$-to-FUV flux ratio is 
also expected to be reproduced when comparing the local H$\alpha$ and FUV fluxes. 
Therefore, a spatially resolved analysis of galaxies will be helpful 
to understand the origin of the deficient H$\alpha$ flux. 

Since the FUV continuum is more affected by dust attenuation than the H$\alpha$ 
emission, corrections for extinction must be properly taken into account to derive 
the intrinsic SFR$_\mathrm{FUV}$ and SFR$_\mathrm{H\alpha}$. 
While the Balmer decrement has been widely used as a direct indicator of the internal
attenuation if the spectral data is available \citep[e.g.,][]{2009ApJ...706..599L,
2012ApJ...750..122B,2015MNRAS.450.3381L,2018A&A...613A..13Y,2019ApJS..242...11L,
2020PASP..132i4101W}, spectral energy distribution (SED) fitting is also widely used as an 
indirect method \citep[e.g.,][]{2004A&A...419..821T,
2018MNRAS.476.1705S,2018A&A...613A..13Y,2021RAA....21....6W,2020arXiv201105918J}.
Indeed, star-forming galaxies have both radial and azimuthal variations in internal attenuation 
\citep[e.g.,][and references therein]{2020MNRAS.495.2305G}. 
Hence, spatially resolved SED fitting with multi-band photometric data has 
the advantage of being able to characterize not only local dust attenuation but also the 
physical properties of the stellar population, which help explore the evolution of galaxies. 

In order to robustly analyze the spatially resolved SF in galaxies, it 
would be efficient to focus on galaxies with extended star-forming disks. 
This would also make it easier to compare the H$\alpha$-to-FUV 
flux ratio with previous studies since star formation in extended disks and dwarf galaxies share 
similar characteristics, such as low surface density and vulnerability to stellar feedback. 
The Galaxy Evolution Explorer (GALEX) satellite \citep{2005ApJ...619L...1M} 
discovered galaxies with UV-bright features in the outskirts 
\citep{2005ApJ...627L..29G,2005ApJ...619L..79T}, revealing that they are not 
rare populations, as they occupy $\sim$20--30 percent of the local disk 
galaxies \citep{2007ApJS..173..538T,2007AJ....134..135Z, 2011ApJ...733...74L}. 
Therefore, these extended UV (XUV) disk galaxies would be the best 
target for analyzing how SF occurs and evolves in galaxy outskirts. 

\begin{figure*}[ht]
\centering
\vspace{3mm}
\includegraphics[width=180mm]{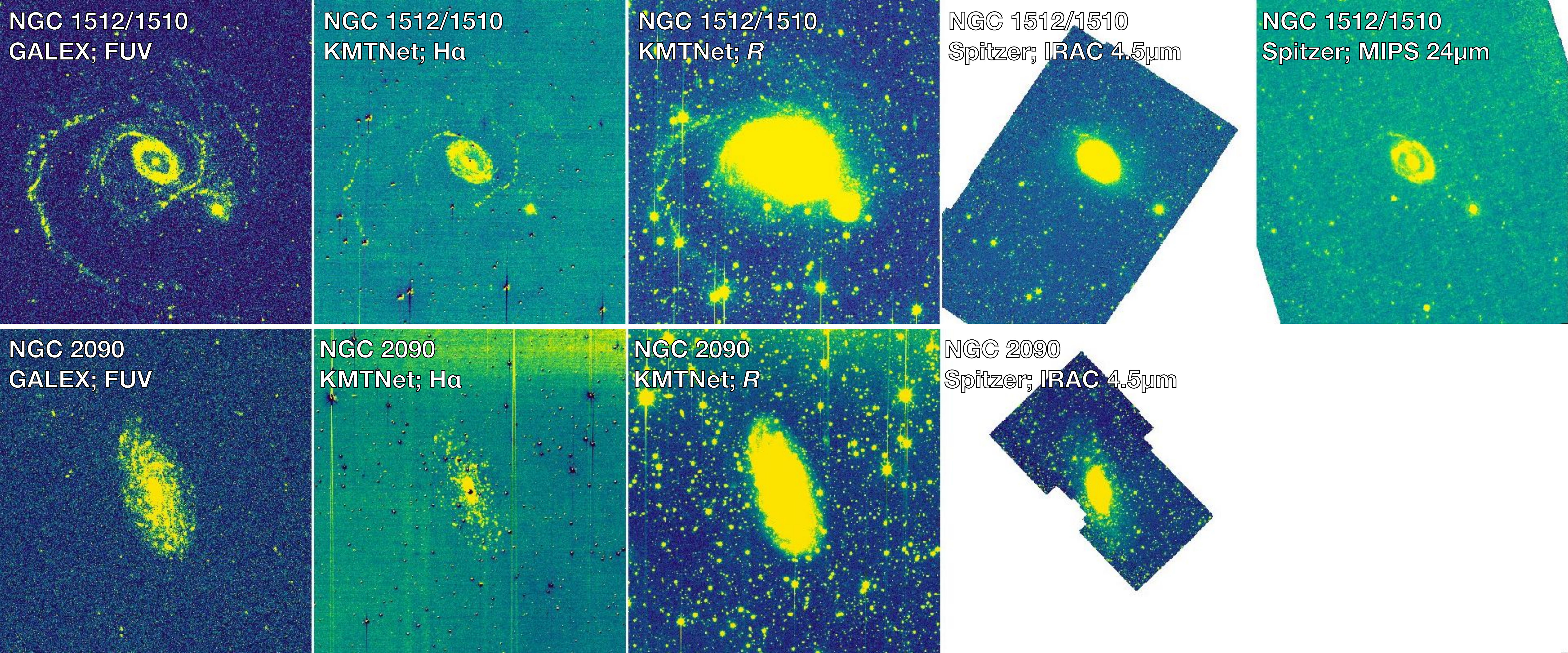}
\caption{Example images of galaxies obtained from GALEX, KMTNet, and Spitzer. 
The top and bottom panels represent NGC 1512/1510 and NGC 2090, respectively. 
The photometric bands are denoted in each panel. The FoV of 
each panel is $\sim$$20\arcmin\times20\arcmin$, corresponding to $\sim$$60\times60$ 
kpc$^2$ at a distance of 10 Mpc. \label{fig:fig1}}
\end{figure*}

In this paper, we investigate spatially resolved SFRs derived from the FUV 
and H$\alpha$ fluxes in the star-forming XUV-disk galaxies. Furthermore, we examined the 
main factors causing a deficient H$\alpha$ flux in the extended disks of galaxies 
using SED fitting. The H$\alpha$ maps of the XUV-disk galaxies were obtained from 
a deep and wide-field imaging survey of nearby galaxies in the southern hemisphere 
(Kim et al. in prep.) obtained with the Korea Microlensing Telescope Network 
\citep[KMTNet;][]{2016JKAS...49...37K}. 

This paper is structured as follows. We describe the photometric data 
obtained from our survey and 
the archival data set utilized for SED fitting in Section \ref{sec:data}. In the 
same section, we summarize the target galaxies, data reduction, and H$\alpha$ flux 
calibration. In Section \ref{sec:method}, we describe the method for measuring fluxes 
in grid-shaped apertures, and the input parameters for SED fitting. The results of the flux 
measurements and the obtained properties of galaxies are presented in Section 
\ref{sec:result}. In Section \ref{sec:discussion}, we discuss possible causes 
of the H$\alpha$ deficit in the extended disks of galaxies, and the 
summary is presented in Section \ref{sec:summary}. 

\section{Observations and data reduction} \label{sec:data}

The KMTNet Nearby Galaxy Survey was started to investigate the 
nature of the low surface brightness regime. 
KMTNet consists of three 1.6 m telescopes located at Cerro Tololo 
Inter-American Observatory (CTIO), South African Astronomical Observatory 
(SAAO), and Siding Spring Observatory (SSO). The pixel scale is 0.4 arcsec 
and the typical seeing is 1.1 arcsec. 
Thanks to the capability of its wide 
field-of-view (FoV $\sim 12$ deg$^2$) and deep photometric depth 
($\mu_{1\sigma}\sim$ 28--29 mag arcsec$^{-2}$), we could detect 
the faint and extended structures of NGC 1291 \citep{2018AJ....156..249B}, 
and discover its dwarf satellite candidates \citep{2020ApJ...891...18B}. 
All of these results were achieved by using co-added images, 
which have an integration time of $\sim$3.0 and $\sim$1.5 hr in the 
$B$- and $R$-band, respectively. 

We have now compiled data sets in the optical broadbands (Johnson-Cousins $BVRI$) 
and H$\alpha$ narrowband, and every single image was taken with an exposure 
time of 120 seconds. The details of the survey will be described in Kim et al. 
(in prep.). In this study, we selected two star-forming galaxies with good photometric 
quality: NGC 1512 and NGC 2090. The observations were carried out between 2017 
January and 2019 December in dark time. Broadband data were taken at KMTNet-CTIO, 
KMTNet-SAAO, and KMTNet-SSO. The observations were allocated to use a single 
band at a single site if possible, to minimize the uncertainty that may arise from 
the subtly different conditions at each site. In addition, we augmented the 
integration time of the $B/R/I$ bands compared to our previous studies, in order to 
enhance the depth of the targets, yielding $\sim$4.8/5.0/4.6 hr for NGC 
1512 and $\sim$4.6/6.3/4.7 hr for NGC 2090, respectively. H$\alpha$ data were taken 
at KMTNet-CTIO with an integration time of $\sim$1.9 hr for both galaxies. Note that 
the central wavelength of the H$\alpha$ narrowband is $\sim$6570\AA\ with a 
bandwidth of $\sim$80\AA, so that the H$\alpha$ emission from our targets 
($z\simeq0.003$) lies within the bandpass of the H$\alpha$ filter. 

Additionally, we utilized archival images covering FUV to MIR. 
Finally, a total of 10--11 data were used for each galaxy, including {\it GALEX} 
FUV/NUV, {\it KMTNet} $B/R/I$/H$\alpha$, {\it Spitzer} IRAC 
3.6/4.5/5.8/8$\mu$m, and {\it Spitzer} MIPS 
24$\mu$m if available. Sample images of the galaxies are shown in 
Figure \ref{fig:fig1}. 

\subsection{The target galaxies} \label{sec:data:sample}

{\it NGC 1512}. This galaxy has prominent features with inner and outer spiral structures. 
In \cite{2007ApJS..173..538T}, it was classified as a Type 1 XUV-disk 
galaxy that 
shows structured UV-bright complexes beyond the SF threshold, 
corresponding to $\mu_\mathrm{FUV}\sim 27.25$ AB mag arcsec$^{-2}$. This 
galaxy has been considered an interacting system with a blue compact 
dwarf galaxy NGC 1510, located in the south-west direction. The disturbed 
spirals in the north-west region and filamentary structure between the two galaxies 
clearly indicate the signs of interaction. \cite{2009MNRAS.400.1749K} reported 
a tight correlation between the dense neutral gas and UV-bright clumps, and 
discovered two probable tidal dwarf galaxies. These indicate that the interaction 
between NGC 1512 and NGC 1510 may have triggered the recent star formation 
activity in NGC 1512. 

\begin{deluxetable}{lcc}[t]
\tablenum{1}
\tablecaption{Galaxy informations \label{tab:tab1}}
\tablewidth{0pt}
\tablehead{
\colhead{} & \colhead{NGC 1512} & \colhead{NGC 2090}
}
\startdata
R.A.\tablenotemark{\scriptsize{a}} (J2000) & 04:03:54.28 & 05:47:01.89 \\
Decl.\tablenotemark{\scriptsize{a}} (J2000) & $-$43:20:55.9 & $-$34:15:02.2 \\
Morphology\tablenotemark{\scriptsize{a}} & SB(r)a & SA(rs)c \\
$m_B$\tablenotemark{\scriptsize{a}} (mag) & 11.13$\pm$0.10 & 11.99$\pm$0.13 \\
Recessional velocity\tablenotemark{\scriptsize{a}} (km s$^{-1}$) & 898$\pm$3 & 921$\pm$2 \\
Distance\tablenotemark{\scriptsize{b}} (Mpc) & 10.4 & 11.3 \\
XUV-disk type\tablenotemark{\scriptsize{b}} & Type 1 & Type 2 \\
$D_{25}$\tablenotemark{\scriptsize{b}} (arcmin) & 7.0 & 4.3 \\
$A_V$\tablenotemark{\scriptsize{c}} (mag) & 0.03 & 0.11\\
\enddata
\tabletypesize{\small}
\tablenotetext{a}{NASA Extragalactic Database (NED)}
\tablenotetext{b}{\cite{2007ApJS..173..538T}}
\tablenotetext{c}{\cite{2011ApJ...737..103S}}
\end{deluxetable}

{\it NGC 2090}. This galaxy might be less fascinating than NGC 1512 as it 
consists of clumpy flocculent spirals. Since it has only minimal UV 
emission beyond the SF threshold, \cite{2007ApJS..173..538T} classified this galaxy as 
a Type 2 XUV-disk galaxy, which has blue FUV$-$NIR color in a large and faint outer disk. 
They also described that NGC 2090 is likely to reside in an isolated environment. 

\cite{2007ApJS..173..538T} claimed that XUV-disk galaxies tend to be 
gas-rich systems, in which Type 1 features may be induced by some external 
perturbations. A spatially resolved analysis of these two types of XUV-disk
galaxies will be crucial to not only inspect the cause of the deficient H$\alpha$ flux in 
the low SFR regime but also provide insight into the mechanism 
triggering the recent SF in the extended disks. The major properties of the galaxies 
are summarized in Table \ref{tab:tab1}. 

\subsection{Data reduction and calibration} \label{sec:data:redux}

The data reduction was carried out in the same way described in 
\cite{2018AJ....156..249B}. In brief, we followed the standard IRAF procedures, 
which include overscan correction and flat-field correction 
using a dark-sky flat. We performed background 
subtraction for individual images using modeled planes with PYTHON. Using 
SExtractor \citep{1996A&AS..117..393B} and SCAMP \citep{2006ASPC..351..112B}, 
we solved astrometric calibrations as instructed by the KMTNet 
team.\footnote{\url{http://kmtnet.kasi.re.kr}} Finally, we created co-added images using 
SWarp \citep{2002ASPC..281..228B}. From the RMS noise per pixel, 
the 1$\sigma$ depths of surface brightness in the optical broadbands were calculated 
as low as $\sim$27--29 mag arcsec$^{-2}$. 

Since the H$\alpha$ narrowband image contains both stellar continuum and nebular 
emission, continuum subtraction is necessary. To subtract the scaled 
$R$-band image from the H$\alpha$ image, we firstly calculated a scale 
factor defined as $\mathrm{BNCR}\equiv c_B/c_N$, where $c_B$ and $c_N$ are 
the fluxes of the field stars measured in the broad- and narrow-band images, respectively. 
Note that most of the field stars were assumed to have no H$\alpha$ emission. 

\begin{figure}[t]
\centering
\includegraphics[width=85mm]{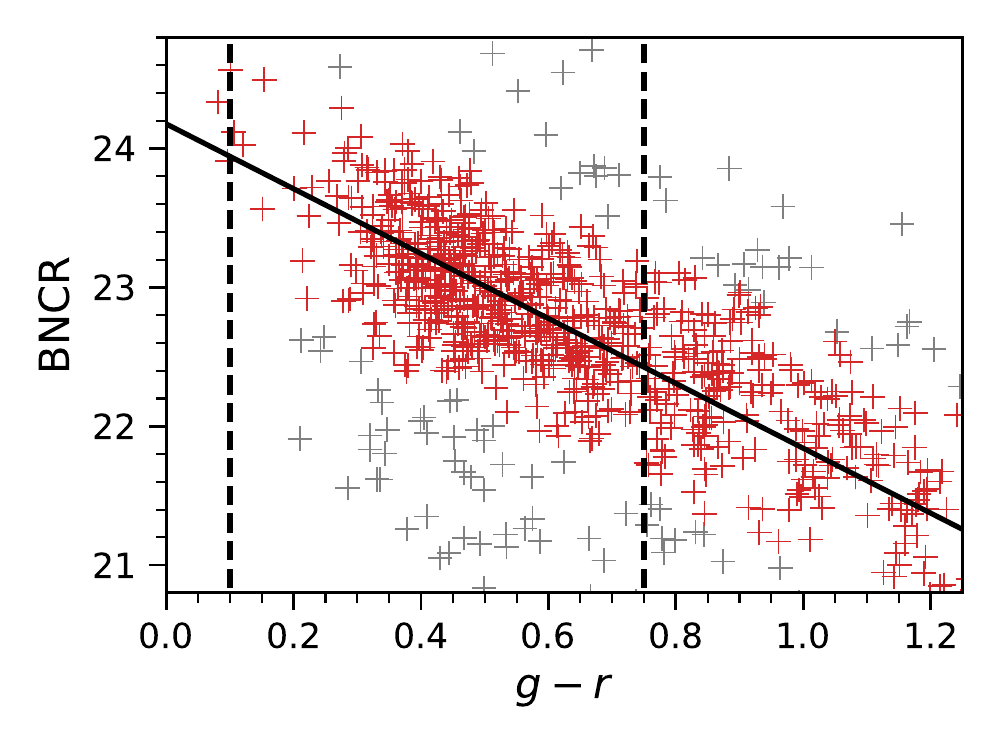}
\caption{Calculated BNCR distribution of field stars in the NGC 1512 image as a function 
of their $g-r$ colors. The red crosses represent the data selected by iterative 
sigma clipping and used for a linear fit (solid line). 
The vertical dashed lines represent 
the minimum and maximum colors of the galaxy, respectively. 
\label{fig:fig2}}
\end{figure}

Figure \ref{fig:fig2} shows the calculated BNCR distribution of NGC 1512 
data as a function of the $g-r$ color of the field stars, which were obtained from 
the AAVSO Photometric All-Sky Survey (APASS) DR9.\footnote{\url{https://www.aavso.org/apass}} 
It reveals that the BNCR depends on the color, showing a large scatter of $\sigma\simeq0.4$. 
We note that NGC 1512 showed a color variation of $B-R\sim0.45$--1.45 mag (see Figure \ref{fig:fig5}), 
corresponding to $g-r\sim$ 0.10--0.75 mag.\footnote{It is transformed by using the equations from 
\url{https://www.sdss.org/dr16/algorithms/sdssubvritransform/\#Lupton(2005)}} 
Using a high BNCR when subtracting the continuum can be risky,
because it can introduce a false H$\alpha$ emission in the regions dominated 
by old stellar populations. Therefore, we adopted the minimum BNCR, 
corresponding to the reddest color of each galaxy, even though it may result in 
an over-subtraction of the stellar continuum. Indeed, this choice mostly raised 
uncertainties in the regions that have lower $g-r$ colors. Systematic uncertainty 
due to the use of a single BNCR will be discussed in Section \ref{sec:disc:BNCR}. 

We estimated the H$\alpha$ flux using a continuum-subtracted H$\alpha$ 
image, adopting the method described in Appendix A of \cite{2008ApJS..178..247K}. 
In brief, we firstly computed the photometric zero point of the $R$-band image, which 
is expressed as
\begin{equation}
\mathrm{ZP}=-2.5\ \mathrm{log}(\lambda^2\frac{f_\lambda}{\mathrm{CR}})-2.397,
\end{equation}
where $f_\lambda$ is the mean flux density of $217.7\times 10^{-11}$ 
erg s$^{-1}$ cm$^{-2}$ \AA$^{-1}$ in the $R$-band \citep{1998A&A...333..231B}, 
and CR is the measured count rate [count s$^{-1}$] of the field stars. 
Then, we calibrated the H$\alpha$ flux using the equations
\begin{equation}
U\equiv\frac{f_\lambda}{\mathrm{CR}}=\lambda^{-2}10^{-0.4(\mathrm{ZP}+2.397)},
\end{equation}
\begin{equation}
f_{cal}(\mathrm{H\alpha)}=U\times \mathrm{FWHM_{H\alpha}\times CR_{H\alpha}\times BNCR},
\end{equation}
where FWHM$_\mathrm{H\alpha}$ is an H$\alpha$ bandwidth of 80\AA, and 
CR$_\mathrm{H\alpha}$ is the measured count rate of the source in the 
continuum-subtracted H$\alpha$ image. Since the photometric zero point 
was computed using the $R$-band image, we multiplied CR$_\mathrm{H\alpha}$ 
by BNCR to adapt to the same zero point. Note that all the single images taken with 
KMTNet have the identical exposure time of 120 sec, so the count normalization by 
exposure time does not affect the above calibration in practice. Finally, transmission 
correction for the KMTNet filter set was applied. The surface 
brightness limit was calculated to be $\sim$$7\times10^{-18}$ erg s$^{-1}$ cm$^{-2}$ arcsec$^{-2}$, 
which is comparable to the depth in \cite{2008ApJS..178..247K}. 
We note that the H$\alpha$ flux in this study is a composite 
of H$\alpha$ and [N\textsc{ii}] lines, so our measurements for the H$\alpha$ flux 
can be considered as an upper limit. 

\section{Method} \label{sec:method}

\subsection{Grid-shaped aperture photometry} \label{sec:method:grid}

\begin{figure}[t]
\centering
\includegraphics[width=85mm]{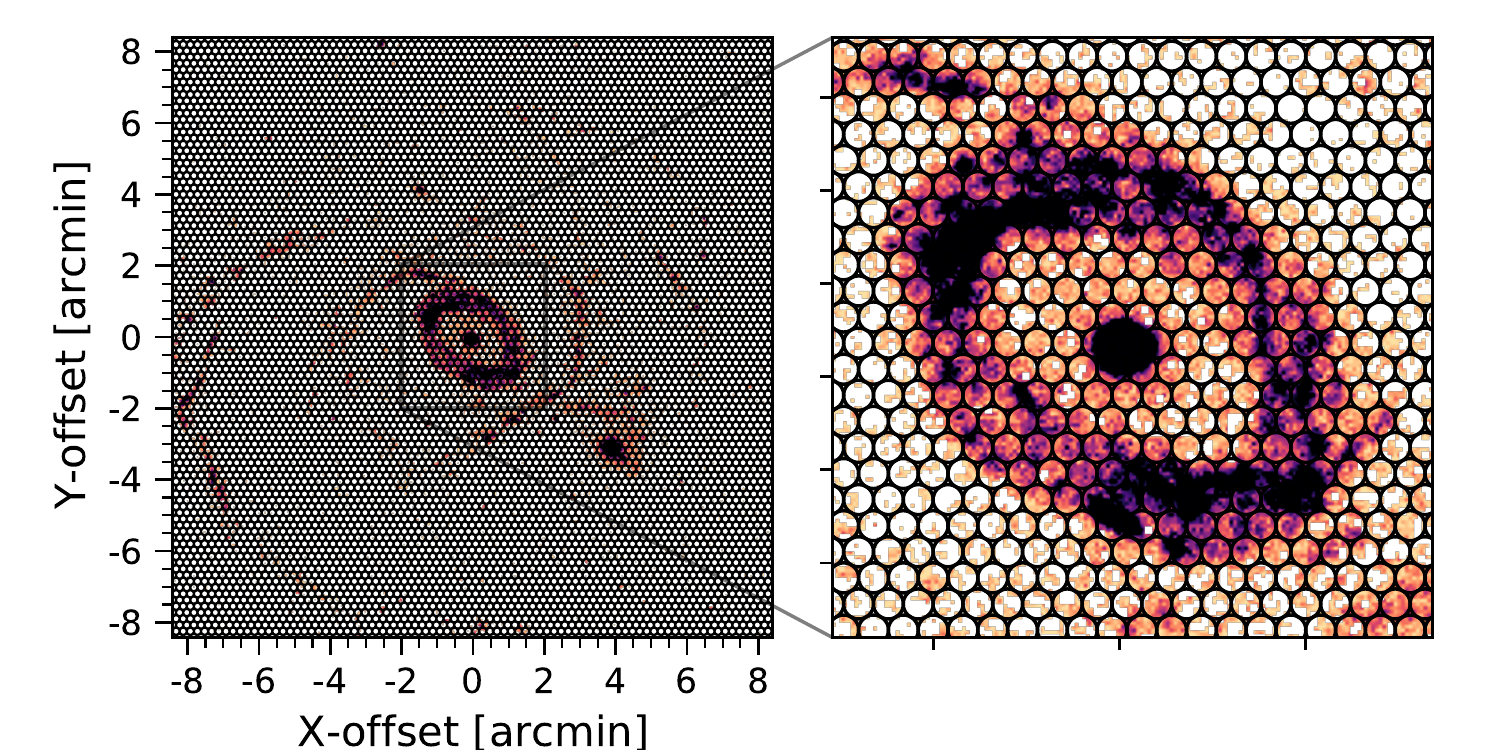}
\caption{Aperture placement for spatially resolved photometry, drawn over the 
FUV image of NGC 1512. All the apertures have the same radius of 6 arcsec 
and are arranged to avoid overlapping each other. 
\label{fig:fig3}}
\end{figure}

Prior to performing the photometry, we carried out a background subtraction 
for all of the FUV-to-MIR data. Then, we placed adjacent circular apertures that 
covered the entire galaxy, as shown in Figure \ref{fig:fig3}. This hexagonal placement 
allowed us to perform spatially resolved photometry while minimizing the loss of flux.
The aperture size was set to be 6 arcsec in radius, corresponding to 300 pc at a 
distance of 10 Mpc. This aperture seemed adequate to cover the physical size 
of typical H\textsc{ii} regions \citep[see][and references therein]{2009A&A...507.1327H}. 

It is worth mentioning that a resolution bias can occur when determining 
properties within apertures in spatially resolved analyses 
\citep[see][]{2018MNRAS.476.1705S,2018MNRAS.476.1532S}. A smaller aperture 
size may introduce an incorrect flux measurement due to contamination by the light of 
neighboring sources. We found that the SED 
fitting results, especially internal attenuation, tended to be inaccurately derived in 
crowded regions when using an aperture size smaller than 6 arcsec. 

Such grid-shaped apertures with a radius of 6 arcsec are quite dense, so it enabled 
significant statistical sampling. Furthermore, it minimized the effect of aperture 
correction. The FWHM of the point spread functions of the GALEX, KMTNet, 
and IRAC data are smaller than 5 arcsec, while that of MIPS data is about 6 
arcsec. We computed the aperture correction factors corresponding to the aperture size of 6 
arcsec by referring to the descriptions of each instrument.\footnote{
\url{http://www.galex.caltech.edu/} and \url{https://irsa.ipac.caltech.edu/}} 
Consequently, the measured fluxes were scaled up by $\lesssim$10\% 
for FUV-to-IRAC 8.0$\mu$m, and $\sim$80\% for MIPS 24$\mu$m. 

We corrected the fluxes for the foreground reddening $A_V$, which was 
derived from the maps of \citet{2011ApJ...737..103S} and the extinction 
law of \citet{1989ApJ...345..245C} with $R_V=3.1$. It yielded the 
correlations of $A_\mathrm{FUV}=2.7A_V$, $A_\mathrm{NUV}=3.2A_V$, 
$A_B=1.3A_V$, $A_{R(\mathrm{H\alpha})}=0.8A_V$, and $A_I=0.5A_V$. 
Indeed, we designed the survey avoiding galaxies in the low Galactic 
latitudes to minimize contamination from the Galactic cirrus, 
so the foreground reddening appears to be weak or 
negligible.\footnote{\url{https://irsa.ipac.caltech.edu/applications/DUST/}} 

Since the image conditions for the two galaxies were slightly different, 
it was hard to select the star-forming regions using a single common 
threshold in FUV (or H$\alpha$) flux. Therefore, we firstly limited the FUV 
fluxes to $\mathrm{log}\ f_\mathrm{FUV}\ge-28.4$ and $-$28.1 erg s$^{-1}$ 
cm$^{-2}$ Hz$^{-1}$ for NGC 1512 and NGC 2090, respectively. Then, we 
empirically eliminated the sky background and non-astronomical objects with 
the combination of H$\alpha$-to-FUV flux ratio, and optical colors. 
In addition, we removed the pixels contaminated by foreground stars and 
saturation trails. Note that the pixels in nuclear regions of the galaxies were 
also discarded to minimize contaminations by spheroidal components and 
latent active galactic nucleus (AGN). As a result, a total of 387 and 377 regions 
were selected as star-forming clumps in NGC 1512 and NGC 
2090, respectively. The error budgets of the flux measurements for star-forming 
regions were computed by taking the RMS within each aperture and 
background scatter into account. 

\begin{figure}[t]
\centering
\includegraphics[width=85mm]{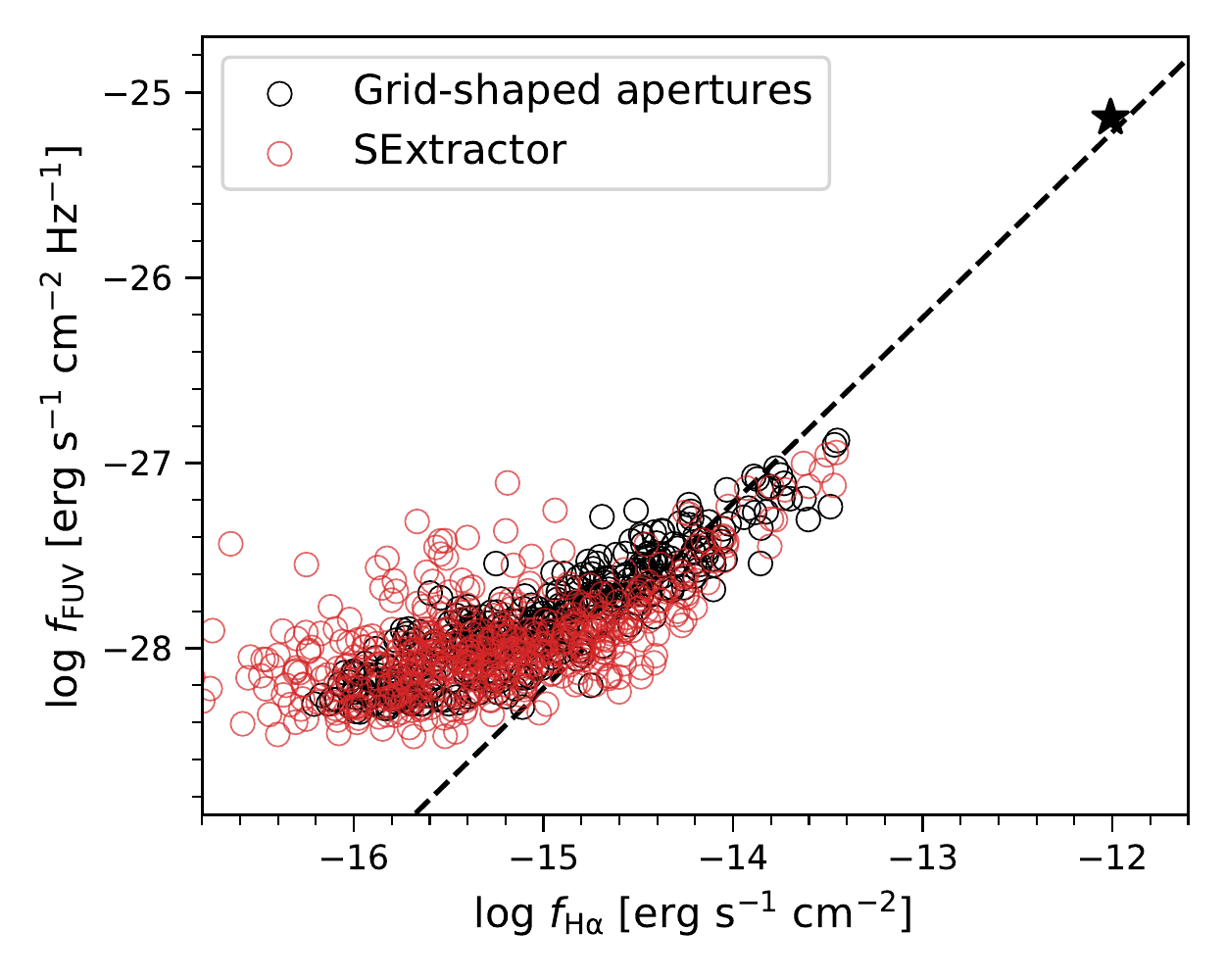}
\caption{Comparison of the FUV and H$\alpha$ fluxes of star-forming regions in 
NGC 1512, measured using grid-shaped apertures (black circles) and SExtractor 
(red circles). There appears to be broadly good agreement between the results. 
Some discrepancy in the low H$\alpha$ flux regions is the result of unmasked 
cavities in the H$\alpha$ image. The dashed line indicates when the SFRs, 
converted following \cite{1998ARA&A..36..189K} with Chabrier IMF, are consistent with 
each other. The black star represents the total flux of 
star-forming regions. \label{fig:fig4}}
\end{figure}

The placement of the grid-shaped apertures is simple and allowed us to 
directly reproduce the properties of galaxies as a 2-D maps. However, some 
star-forming clumps may not be exactly centered in the aperture, introducing an incorrect 
flux measurement. Meanwhile, SExtractor is a widely used tool, which finds the 
center of the source and measures the flux. In order to assess whether the 
grid-shaped apertures measured the fluxes correctly, we independently carried out photometry 
using SExtractor, and compared the measured fluxes. The aperture 
size was fixed to the same radius of 6 arcsec, using FLUX\_APER in SExtractor. 

Figure \ref{fig:fig4} shows the FUV and H$\alpha$ flux distributions of the 
star-forming regions in NGC 1512 measured using the two photometry 
methods. The result of grid-shaped aperture photometry was found to be 
broadly consistent with the result from SExtractor, even though there is 
no guarantee that the samples were identical to each other. This implies 
that there was no critical defect in the flux measurement with the grid-shaped 
aperture photometry. Note that there were some deviations between the results 
derived by the two methods in regions of low H$\alpha$ flux. This mainly 
originated from cavities in the H$\alpha$ image created 
when the continuum of saturated stars was subtracted. 

Interestingly, the H$\alpha$-to-FUV flux ratio decreased as the flux decreased. 
This finding is in good agreement with previous studies 
\citep[e.g.,][]{2009ApJ...706..599L} that showed a deficient 
H$\alpha$ flux in low luminosity systems such as dwarf galaxies. The only 
difference is that we are dealing with the spatially resolved fluxes for a single 
galaxy. We estimated the total fluxes by summing the fluxes of the star-forming 
regions, shown as a black star in Figure \ref{fig:fig4}. If we had estimated the total 
flux of such a galaxy, a deficient H$\alpha$ flux would not have been discovered. 

\subsection{SED modeling} \label{sec:method:sed}

\begin{deluxetable}{lc}[t]
\tabletypesize{\scriptsize}
\tablenum{2}
\tablecaption{Parameters used in the SED fitting \label{tab:tab2}}
\tablewidth{0pt}
\tablehead{
\colhead{Parameter} & \colhead{Value}
}
\startdata
\multicolumn{2}{c}{Single Stellar Population (SSP)} \\
\cline{1-2}
Model & \cite{2003MNRAS.344.1000B} \\
Initial mass function & \cite{2003PASP..115..763C} \\
Metallicity ($Z$) & 0.02 \\
Separation age\tablenotemark{\scriptsize{a}} & 10 Myr\\
\cline{1-2}
\multicolumn{2}{c}{Nebular emission} \\
\cline{1-2}
Ionization parameter & log$U=-2$ \\
LyC escape fraction & 0.0, 0.1, 0.2, 0.3 \\
\cline{1-2}
\multicolumn{2}{c}{``Delayed" double-exponential SFH} \\
\cline{1-2}
age$_\mathrm{main}$ & 13 Gyr\\
$\tau_\mathrm{main}$ & 1, 3, 5, 7 Gyr\\
$f_\mathrm{burst}\tablenotemark{\scriptsize{b}}$ & 0.001, 0.002, 0.005, 0.01, 0.02, 0.05, 0.1 \\
age$_\mathrm{burst}$ & 5, 15, 45, 130, 400 Myr\\
$\tau_\mathrm{burst}$ & 1, 5, 15, 50 Myr\\
\cline{1-2}
\multicolumn{2}{c}{Dust attenuation} \\
\cline{1-2}
Name & \cite{2000ApJ...539..718C} \\
$A_V^\mathrm{ISM}$ & 0.01, 0.1, 0.2, 0.3, 0.4, 0.5, 0.6, 0.7, 0.8, 0.9, 1.0 \\
$\mu$\tablenotemark{\scriptsize{c}} & 0.44, 0.58, 0.72, 0.86, 1.00 \\
Power law slope\tablenotemark{\scriptsize{d}} & $-0.7$ (ISM), $-1.3$ (BC) \\
\cline{1-2}
\multicolumn{2}{c}{Dust emission} \\
\cline{1-2}
Dust template & \cite{2014ApJ...784...83D} \\
$\alpha$ & 0.0625, 1.3125, 2.6250, 4.0000
\enddata
\tabletypesize{\small}
\tablenotetext{a}{Criterion for dividing young and old stars}
\tablenotetext{b}{Mass ratio between the late burst and main populations}
\tablenotetext{c}{$A_V^\mathrm{ISM}/(A_V^\mathrm{ISM}+A_V^\mathrm{BC})$, see details in Section \ref{sec:disc:atten}.}
\tablenotetext{d}{Index $m$ of the optical depth ($\tau_\lambda\propto\lambda^{m}$) in the ambient interstellar medium (ISM) and birth clouds (BC).}
\end{deluxetable}

\begin{figure*}[t]
\centering
\includegraphics[width=180mm]{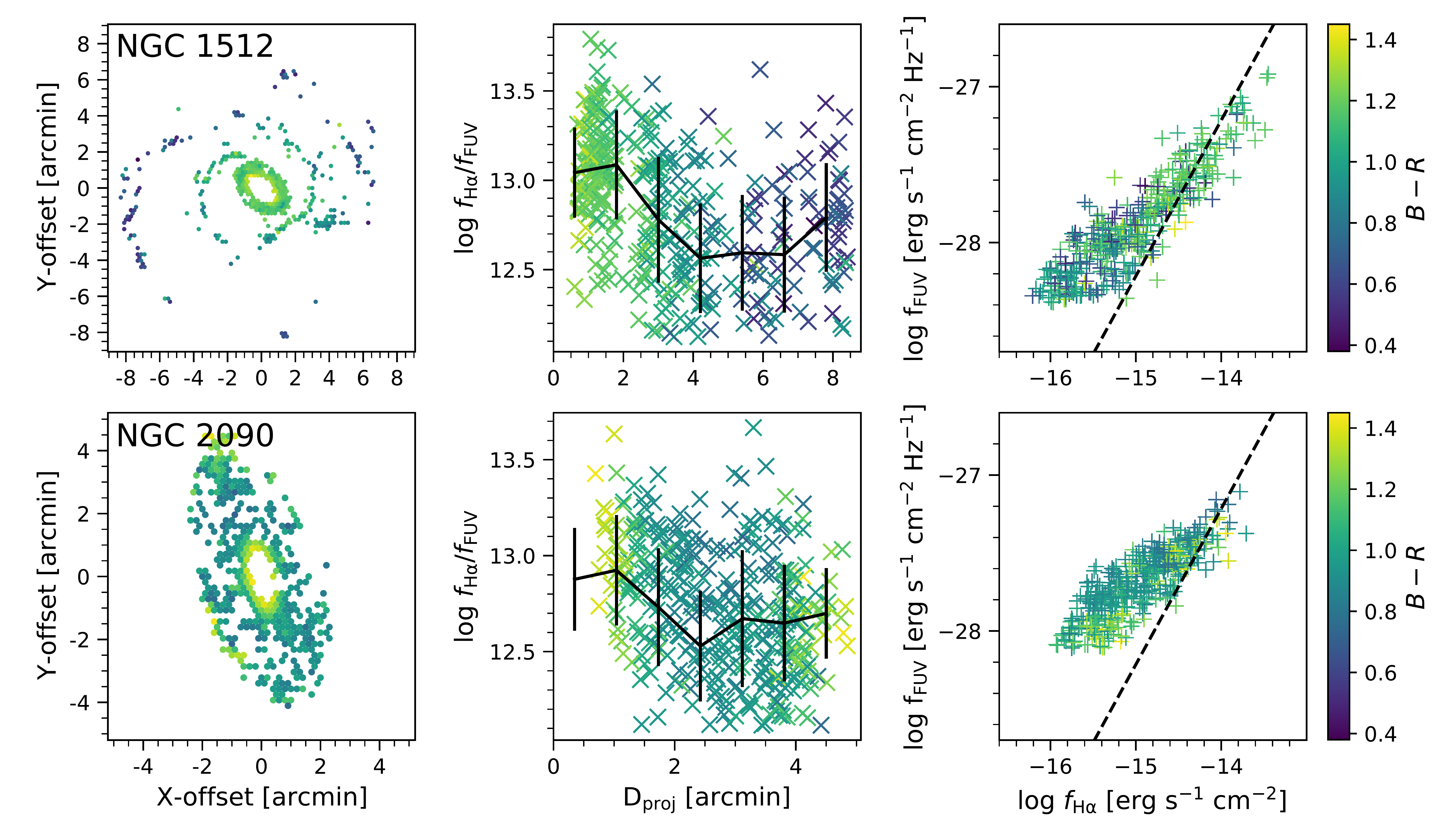}
\caption{Measured photometric properties of NGC 1512 (top) and NGC 2090 
(bottom). All the panels are color-coded by $B-R$ color in the same range.
Left: 2-D maps of selected star-forming regions in the galaxies. There 
is a negative radial gradient in NGC 1512, but no clear trend is found in NGC 2090. 
Middle: the H$\alpha$-to-FUV flux ratios as a function of projected distances from 
the center of each galaxy. The black lines represent the median values 
in each bin together with the standard deviations. It shows a slight correlation 
between the two, with large scatters. Right: the comparison of FUV 
and H$\alpha$ fluxes. The dashed lines indicate when the SFRs, converted following 
\cite{1998ARA&A..36..189K} with Chabrier IMF, are consistent with each other. A 
deficit of H$\alpha$ flux is prominent where H$\alpha$ flux is low. NGC 1510 was 
excluded in the top panels for clarification.
\label{fig:fig5}}
\end{figure*}

We performed a spatially resolved SED fitting with the Code Investigating GALaxy 
Emission \citep[CIGALE;][]{2019A&A...622A.103B} in order to investigate the 
local properties of individual star-forming regions, such as their internal 
attenuations and star formation histories (SFHs). This software contains various 
modules, which permit detailed adjustments to create SED models, and perform 
an energy balanced fitting with dust absorption in the shorter wavelengths and 
re-emission in longer wavelengths, simultaneously. Note that many SED modeling 
codes have been developed, such as MAGPHYS \citep{2008MNRAS.388.1595D}. 
We expect that the result is unlikely to significantly change regardless of which code 
we use \citep[see][]{2019A&A...621A..51H}. 

\begin{figure*}[t]
\centering
\includegraphics[width=180mm]{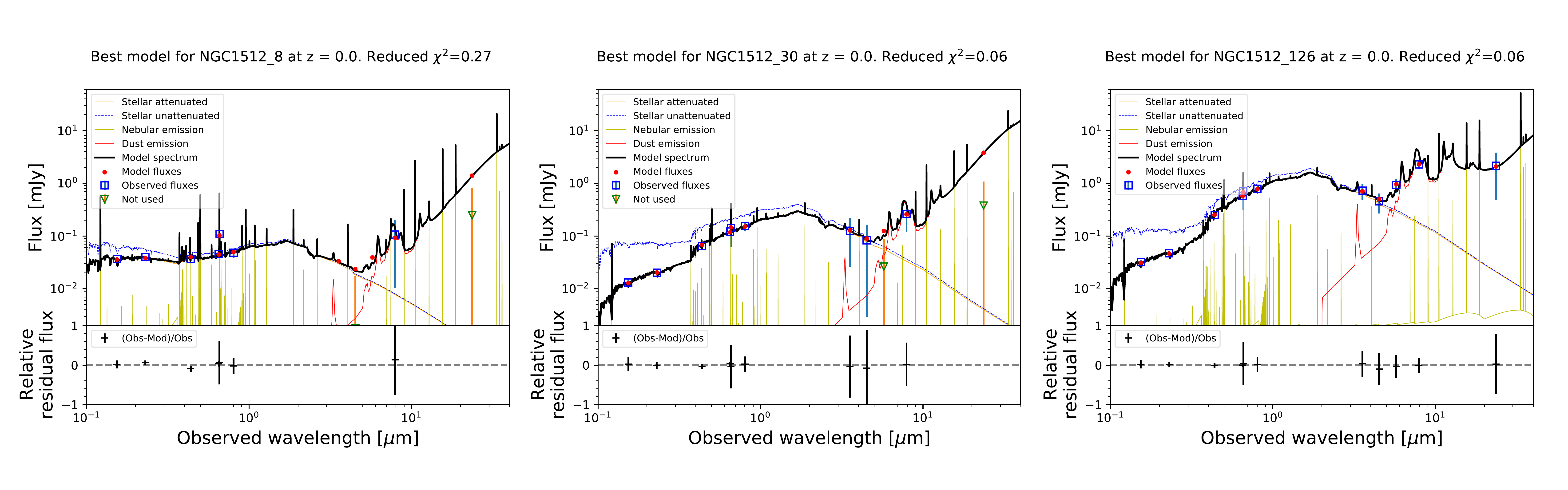}
\caption{Examples of modeled SED for three representative regions in NGC 
1512: the outer spirals (left), inner spirals (middle), and central regions (right). 
The configurations are based on CIGALE default, except for the inverse 
triangles, which represent data not used for the fitting due to large uncertainties. 
\label{fig:fig6}}
\end{figure*}

To avoid a waste of computing time by excessive iteration when creating SED 
models, we constrained the optimized parameter ranges, yielding a total of 
$\sim$500,000 models. We adopted the single stellar population model 
of BC03 \citep{2003MNRAS.344.1000B} with Chabrier IMF 
\citep{2003PASP..115..763C}. The metallicity was fixed to $Z=0.02$, which is the 
same as the solar metallicity. Note that this choice did not significantly affect the 
fitting result for the extended disks, which are expected to have lower metallicities 
(see details in Section \ref{sec:disc:metal}). We allowed the Lyman continuum 
(LyC) escape fraction in the range of $f_{esc}=0.0$--0.3, which was recently 
suggested as a feasible boundary condition in individual star-forming clumps 
\citep[see][]{2020ApJ...902...54C,2021arXiv210315854O}. We selected a 
delayed double-exponential SFH model, which consists of the main (old) stellar 
population and late starburst population. Since this study is focused on recent 
SF events, we assigned the main population to have relatively simple SFHs with 
the oldest stellar age of 13 Gyr and e-folding times with large sampling intervals (1, 3, 5, and 7 Gyr). 
On the other hand, the late burst population was allowed to have more various 
histories, which were somewhat biased to the timescales of H$\alpha$ and FUV 
radiations. We adopted the dust attenuation model of \cite{2000ApJ...539..718C}, 
which considers the environment of the interstellar medium (ISM) and birth clouds (BC). 
These were separately applied to stars older and younger than 10 Myr, respectively. 
In addition, based on many observational studies \citep[e.g.,][]{1997AIPC..408..403C,
2011ApJ...738..106W,2006ApJ...647..128E,2017MNRAS.467..239B}, we assumed 
that the attenuation ratio between ISM and BC environments can vary within a range 
of $\mu=0.44$--1.00. Note that the attenuation for H$\alpha$ emission cannot be 
directly estimated from the CIGALE fitting since we used the total flux of H$\alpha$ narrowband for the fit. Instead, we computed it using the 
empirical relation between the attenuation in stellar continuum and that in nebular 
emission (see details in Section \ref{sec:disc:atten}). The dust emission model 
of \cite{2014ApJ...784...83D} was adopted, and no AGN effect was considered. 
All the parameters used in the SED fitting are summarized in Table \ref{tab:tab2}. 

It is worth mentioning that the CIGALE fitting provides average SFRs over the last 10 
and 100 Myr (SFR$_{10}$ and SFR$_{100}$). However, we did not use them for any 
SFR analyses, including a direct comparison to SFRs derived from H$\alpha$ 
and FUV fluxes, because there is no guarantee that the entire H$\alpha$ and FUV fluxes are 
originated from the stars exactly formed over the last 10 and 100 Myr, 
respectively. Instead, we used them to investigate the existence of late bursts (see Section 
\ref{sec:disc:sfh}). 

\section{Results} \label{sec:result}

\subsection{Photometric properties} \label{sec:result:obs}

Even though each flux might be affected by the internal characteristics of 
the star-forming regions, it is meaningful to 
examine the photometric properties before analyzing the SED fitting result. 
Note that NGC 1510, which was not our 
interest, is not shown in the figures hereafter. 

Figure \ref{fig:fig5} shows several properties of the star-forming regions in two galaxies 
measured with the grid-shaped aperture photometry. It reveals that the selection criteria 
for the star-forming regions were appropriate, given that the structures of the galaxies were 
well reproduced. While a portion of the data in the disk of NGC 2090 was discarded, 
it still seems enough to investigate the galaxy's properties. 

The left panels in Figure \ref{fig:fig5} show the spatial distributions of the $B-R$ 
colors in galaxies. NGC 1512 shows a negative radial color gradient as expected from 
the existence of UV-bright extended spirals, which are likely to be less affected by dust 
attenuation. On the other hand, NGC 2090 shows a nearly constant color distribution in 
the disk, with an average of $B-R\sim0.95$ mag. 

We examined the correlation between H$\alpha$-to-FUV flux ratio and radial 
distance as shown in the middle panels of Figure \ref{fig:fig5}. Note that NGC 
2090 has large ellipticity, so we converted the projected distance into major 
axis distance using an axis ratio of $a/b=2$. In NGC 1512, the H$\alpha$-to-FUV 
flux ratio appears to first decrease on average with the radial distance and then 
increase later. In contrast, NGC 2090 shows a slight correlation between the flux 
ratio and radial distance. However, both galaxies show large scatters in the 
H$\alpha$-to-FUV flux ratio at a given radial distance. This indicates that the 
individual star-forming regions may have various characteristics regardless of 
their radial distance from the center of the galaxies. The central regions of the 
galaxies tend to have slightly higher H$\alpha$-to-FUV flux ratios on average 
compared to the areas beyond, possibly due to higher dust attenuation and/or 
[N\textsc{ii}] contamination. 

The right panels in Figure \ref{fig:fig5} show a comparison of the FUV 
and H$\alpha$ fluxes in star-forming regions. As mentioned in Section 
\ref{sec:method:grid}, the H$\alpha$-to-FUV flux ratio decreases with decreasing flux. 
Although the fluxes were measured by using apertures of a similar physical size, 
the flux distributions of the galaxies appear to be different. 
This may be explained by two causes: inherently different SFRs and/or diverse 
contaminations including dust attenuation.
In the next section, we will attempt to present more straightforward comparisons 
using the converted SFRs, which were corrected for several factors 
obtained from the SED fitting. 

\subsection{Comparison of SFR$_{H\alpha}$ and SFR$_{FUV}$} \label{sec:result:sfr}

The spatially resolved SED fitting of the two galaxies worked reasonably 
well, with the result showing the reduced $\chi^2$ in almost all regions estimated 
to be less than 5, with a median of $\sim$0.5. Figure \ref{fig:fig6} shows examples 
of the best-fit SEDs for three representative regions in NGC 1512: the outer spirals, 
inner spirals, and central regions. We note that the SEDs in the inner and outer 
spirals are not fitted well at long wavelengths. Indeed, the IR data in most regions 
of the extended disks were not available due to the signal-to-noise ratio lower than 3 
or the limited FoV of the utilized data set (Figure \ref{fig:fig1}). Meanwhile, the outer 
regions of the spiral galaxies are expected to exhibit relatively weak dust emissions 
\citep[see][]{2017A&A...605A..18C}, implying there are insignificant attenuations. 
Therefore, we expect that we do obtain reasonable results even though SED 
fitting for the extended disks was mainly performed with stellar continuum only. 

\begin{figure}[t]
\centering
\includegraphics[width=83mm]{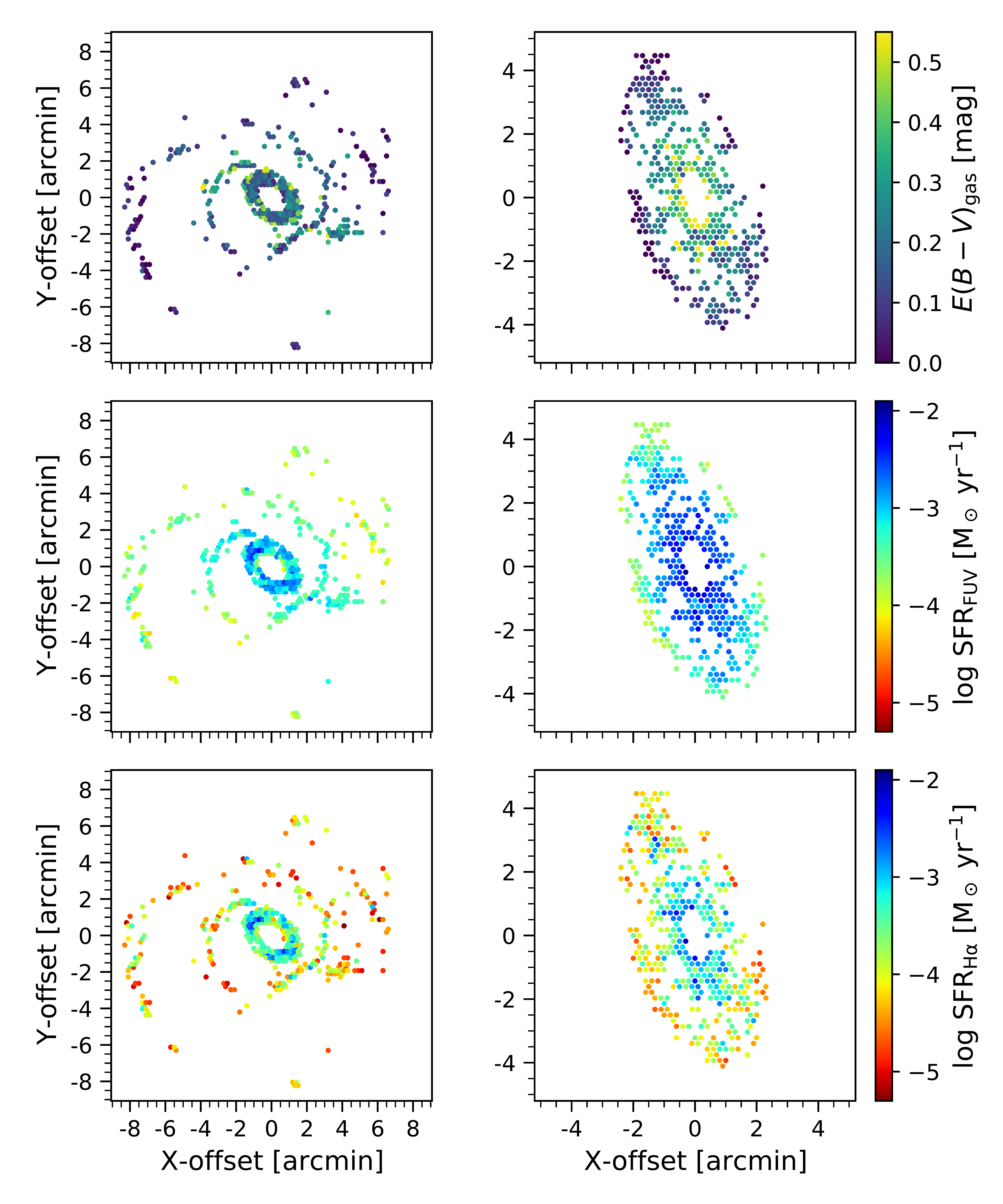}
\caption{Examples of physical properties of NGC 1512 (left) and NGC 2090 (right). 
Each panel is color-coded for each property in the same range. The top panels 
show the 2-D maps of $E(B-V)_\mathrm{gas}$, which is roughly estimated to 
be less than 0.3 mag for most regions. The middle and bottom panels show the SFR 
maps derived from the FUV and H$\alpha$ fluxes, which are corrected for the internal 
attenuation and LyC escape. The disagreement between 
the SFR$_\mathrm{FUV}$ and SFR$_\mathrm{H\alpha}$ evidently exists. 
\label{fig:fig7}}
\end{figure}

In order to estimate the inherent SFRs, we corrected the H$\alpha$ and 
FUV fluxes for the internal attenuation derived from SED fitting. The upper 
panels in Figure \ref{fig:fig7} 
show the 2-D distribution of $E(B-V)_\mathrm{gas}$, corresponding to 
$E(B-V)_\mathrm{BC}$, as will be discussed in Section \ref{sec:disc:atten}. 
These were determined by using $A_V^\mathrm{ISM}$ 
and an attenuation ratio between ISM and BC with an assumption of $R_V=3.1$. Overall, 
$E(B-V)_\mathrm{gas}$ were revealed to be lower than 0.3 mag in most regions 
of both galaxies, while NGC 2090 has a median value slightly higher than 
NGC 1512. For NGC 1512, we confirmed that the results were broadly consistent with the 
results of the spectroscopic analyses in 
\cite{2012ApJ...750..122B} and \cite{2015MNRAS.450.3381L}. 
Furthermore, these results are also similar to that of another 
XUV-disk galaxy, M83 \citep{2007ApJ...661..115G}, indicating that our estimates are reliable. 

We calculated the SFRs using the equations in \cite{1998ARA&A..36..189K} as follows
\begin{equation}
\small
\mathrm{SFR\ [M_\odot\ yr^{-1}]}=7.9\times 10^{-42}\ L_\mathrm{H\alpha}\ \mathrm{[erg\ s^{-1}]}
\end{equation}
\begin{equation}
\small
\mathrm{SFR\ [M_\odot\ yr^{-1}]}=1.4\times 10^{-28}\ L_\mathrm{FUV}\ \mathrm{[erg\ s^{-1}\ Hz^{-1}]}
\end{equation}
with the conversion factors from Salpeter IMF \citep{1955ApJ...121..161S} to 
Chabrier IMF \citep{2003PASP..115..763C} found in \cite{2012ARA&A..50..531K}. 
The H$\alpha$ and FUV fluxes were corrected for the attenuations derived from 
the SED fitting. We assumed that the ionizing photon escape fraction of each 
star-forming clump can reach up to 30\% in the SED fit, and accordingly 
the loss in H$\alpha$ flux was also recovered by using $L_\mathrm{H\alpha}^\mathrm{int} = 
L_\mathrm{H\alpha}^\mathrm{corr}/(1-f_{esc})$, where $L_\mathrm{H\alpha}^\mathrm{int}$ 
and $L_\mathrm{H\alpha}^\mathrm{corr}$ are the intrinsic and attenuation-corrected 
H$\alpha$ luminosities, respectively. 

\begin{figure}[t]
\centering
\includegraphics[width=83mm]{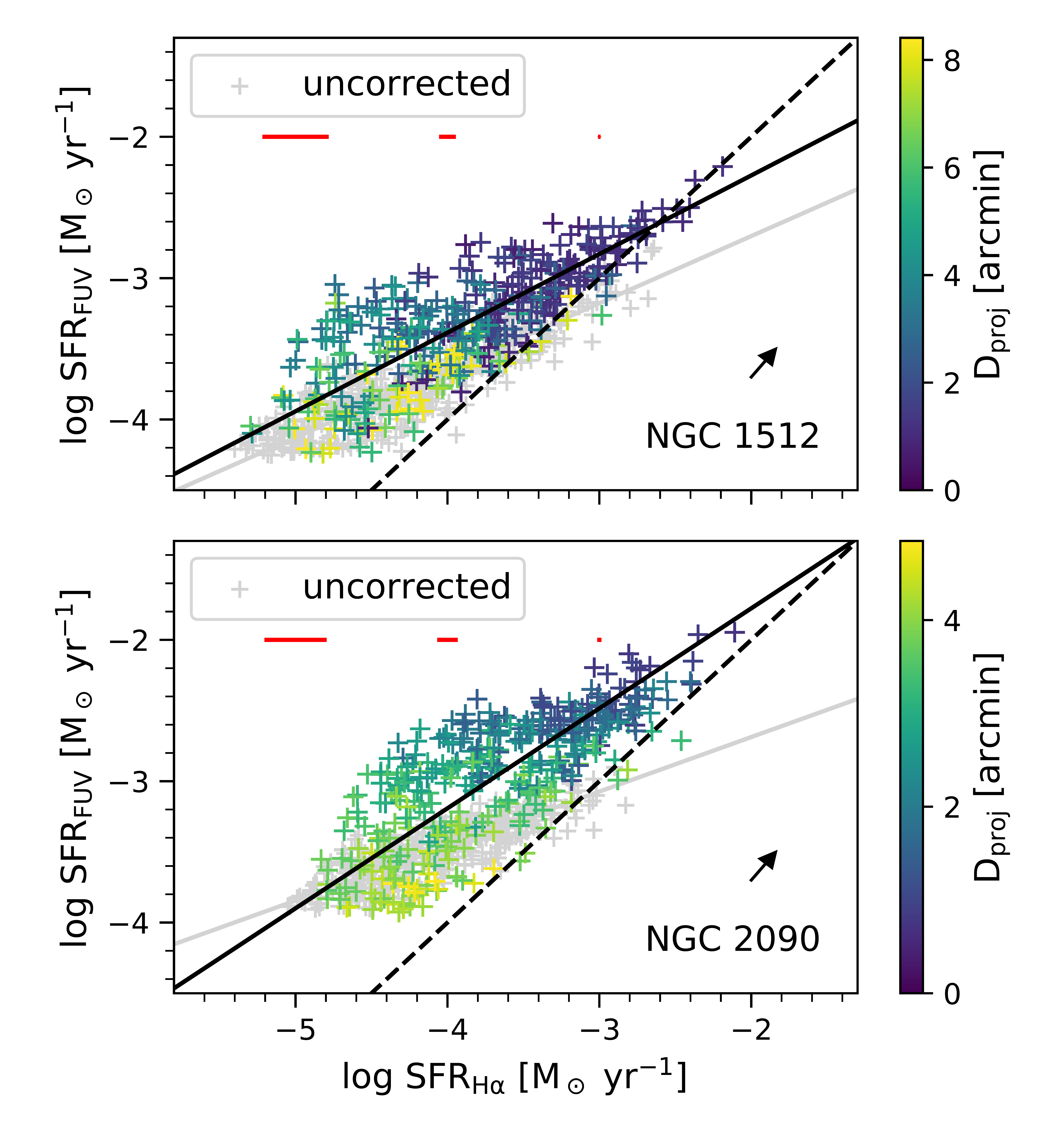}
\caption{Comparisons of the SFR$_\mathrm{FUV}$ and 
SFR$_\mathrm{H\alpha}$ of star-forming regions in NGC 1512 (top) and NGC 
2090 (bottom). The data are color-coded by their radial distances. The solid lines 
represent the results of a linear fit, and the dashed lines 
indicate a one-to-one relation between the SFRs. The red bars displayed at the top 
of each panel represent the mean uncertainties derived from the photometric 
errors for the data at 
$\mathrm{log\ SFR}_\mathrm{H\alpha}=-5\pm0.5$, $-4\pm0.5$, $-3\pm0.5$, 
respectively. The uncertainties for SFR$_\mathrm{FUV}$ are too small to be seen here. 
The small arrow in each panel represents the systematic effect due to the 
choice of IMF (from Chabrier to Salpeter). The data uncorrected for the internal 
attenuation and LyC escape are shown in gray color. 
\label{fig:fig8}}
\end{figure}

The middle and bottom panels in Figure \ref{fig:fig7} show the SFRs derived 
from the FUV and H$\alpha$ fluxes after the corrections in attenuation and 
LyC escape. The two SFRs for most regions of the galaxies appear to be 
clearly different, in the sense that the SFR$_\mathrm{H\alpha}$ tends to be 
systematically lower than SFR$_\mathrm{FUV}$. 

As shown in Figure \ref{fig:fig8}, we compared the SFRs with each other to 
highlight the correlations between them. It is obvious that both galaxies show 
a decline in SFR$_\mathrm{H\alpha}$/SFR$_\mathrm{FUV}$ as the SFR 
decreases. Note that the choice of IMF hardly affect the results, since all the 
points move in the same direction, by as much as 
0.2 dex. We also denoted galactocentric radial distance in color to reveal 
the radial dependence of the discrepancy. While the correlation between radial 
distance and the discrepancy in SFRs is not entirely clear, it can be roughly 
divided into two regions. The central regions of galaxies are distinguishable 
from the rest, as they have higher SFRs and tend to deviate less from 
a one-to-one relation between the SFRs. In contrast, the outer regions exhibit 
lower SFRs and SFR$_\mathrm{H\alpha}$/SFR$_\mathrm{FUV}$, and large 
deviations between SFRs, on average. However, interestingly, the outermost 
regions (light green) deviate less from a one-to-one relation compared to the 
adjacent inner regions (dark green), resulting in significant scatters. The possible 
cause of this trend will be discussed in Section \ref{sec:disc:sfh}. 

\begin{figure}[t]
\centering
\includegraphics[width=83mm]{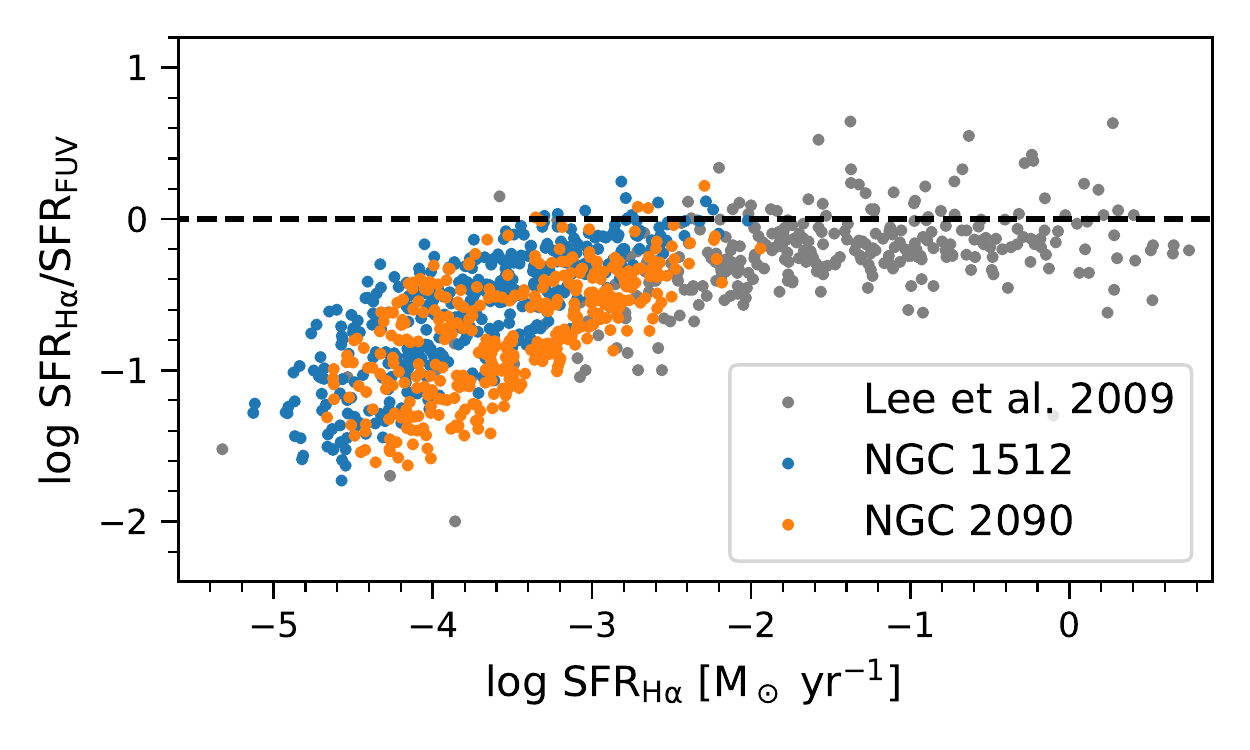}
\caption{Ratio between SFR$_\mathrm{H\alpha}$ and 
SFR$_\mathrm{FUV}$ as a function of the SFR$_\mathrm{H\alpha}$. 
The dashed line indicates the ratio of consistent SFRs. The blue and 
orange points represent the star-forming regions in NGC 1512 and 
NGC 2090, respectively. The gray points are adopted from \cite{2009ApJ...706..599L}.
\label{fig:fig9}}
\end{figure}

Figure \ref{fig:fig9} shows the ratio of SFR$_\mathrm{H\alpha}$ 
to SFR$_\mathrm{FUV}$ as a function of the SFR$_\mathrm{H\alpha}$. 
These were converted into SFRs corresponding to Salpeter IMF in 
order to directly compare with the results of \cite{2009ApJ...706..599L}. 
They claimed that the discrepancy between the SFRs became significant 
at SFR$_\mathrm{H\alpha} \lesssim0.003$ M$_\odot$ yr$^{-1}$. 
Our results extend this finding to a much lower SFR regime. 
Consequently, a deficient H$\alpha$ flux intrinsically exists in low 
SFR regimes. The possible causes of the H$\alpha$ deficit will be 
discussed in the following section.

\section{Discussion} \label{sec:discussion}

We carried out a spatially resolved examination of two XUV-disk galaxies, 
and estimated SFR$_\mathrm{H\alpha}$ and SFR$_\mathrm{FUV}$ in 
each star-forming regions using the SFR conversion recipes in 
\cite{1998ARA&A..36..189K} with Salpeter IMF. Note that these recipes were 
calculated using stellar population models with solar abundance, and each 
calibration for H$\alpha$ and FUV assumed that the SFH is constant for at least 
the past $\sim$10 Myr and $\sim$100 Myr, respectively. 

We found that the deficiency of H$\alpha$ flux found in dwarf galaxies was 
similarly reproduced in the extended disks of galaxies. In other words, the 
discrepancy between SFR$_\mathrm{H\alpha}$ and SFR$_\mathrm{FUV}$ 
appeared to be correlated with the \textit{local} SFR itself. For example, a 
star-forming clump with SFR$_\mathrm{H\alpha}<10^{-4}$ M$_\odot$ yr$^{-1}$ 
shows the deviation between SFRs more than 0.8 dex on average. In this section, 
we will investigate what triggers a deficient H$\alpha$ flux in the extended disks 
of galaxies. 

One of the advantages of SED fitting is that we can restrict each of the involved 
parameters. As introduced in Section \ref{sec:intro}, several possible factors have 
been suggested that can affect SFR$_\mathrm{H\alpha}$/SFR$_\mathrm{FUV}$: 
(1) internal attenuation, (2) metallicity, (3) LyC escape, (4) SFH, and (5) stochasticity 
in high-mass SF. We discuss how significantly the former four factors affect the 
estimate of H$\alpha$ flux using CIGALE. Note that the last factor was not 
considered here, but we clarify that conducting photometry using small apertures 
for individual star-forming clumps with low SFR might have a high probability of 
missing very massive stars. 

\subsection{Internal attenuation} \label{sec:disc:atten}

To estimate the intrinsic SFR using a specific flux, the internal attenuation for 
that wavelength must be corrected properly. From the CIGALE results, we could 
obtain the attenuation for FUV, but not H$\alpha$. Therefore, we calculated the 
attenuation for H$\alpha$ as follows. 

The classical ratio between the two attenuations measured by stellar 
continuum and nebular emission is $\mu=0.44$ \citep{1997AIPC..408..403C,
2011ApJ...738..106W}. However, it has been recently reported that the ratio in highly 
star-forming galaxies can be higher, up to 1 \citep[e.g.,][]{2006ApJ...647..128E,
2010ApJ...712.1070R,2013ApJ...777L...8K,2015ApJ...806..259R,
2015ApJ...807..141P,2017MNRAS.467..239B}. Here, we can regard the 
attenuations for stellar continuum and nebular emission as the attenuations 
in ISM and BC, respectively \citep[see][]{2018A&A...613A..13Y}. Since we 
allowed various ISM-to-BC attenuation ratios between 0.44 and 1 in the SED fitting, 
the attenuation for H$\alpha$ was calculated by using the equation 
$A_\mathrm{H\alpha}=A_R/\mu$ for each star-forming region. As a result, 
$A_\mathrm{FUV}/A_\mathrm{H\alpha}$ varied from 1.2 to 3.0, 
while this ratio in the extended disks tended to be higher than 
that in central regions of galaxies. 

Alternatively, one of the simplest ways to estimate the attenuation for H$\alpha$ 
is using the correlation of $A_\mathrm{FUV}/A_\mathrm{H\alpha}=1.8$, which 
is expected from the Calzetti extinction curve and differential extinction law 
\citep{2001PASP..113.1449C}. Although it is apparently different from the ratio 
calculated above, it would be instructive to check the difference in the resultant 
SFR$_\mathrm{H\alpha}$. Consequently, we found that there is good consistency 
between the two SFR$_\mathrm{H\alpha}$ obtained with different 
$A_\mathrm{H\alpha}$, showing a scatter of $\sim$0.06 dex. This result indicates 
that the deficiency of H$\alpha$ flux in the extended disks of galaxies is unlikely 
to be caused by the incorrect estimation of internal attenuation. 

\subsection{Metallicity} \label{sec:disc:metal}

When we allowed metallicity to be free, the SED fitting results appeared to be 
biased to high metallicity values. Therefore, we fixed metallicity to $Z=0.02$ 
when creating the SED models for the sake of simplicity. This seemed a proper 
assumption since the SFR conversion 
recipes in \cite{1998ARA&A..36..189K} assume a solar abundance with continuous 
SFH. However, the outer regions of late-type galaxies are 
expected to have lower metallicity than the inner regions due to the low efficiency of 
star-forming activity. Indeed, a negative radial 
gradient of metallicity has been observed in NGC 1512 via spectroscopic studies 
\citep[e.g.,][]{2015MNRAS.450.3381L}. Furthermore, the XUV-disk galaxies 
have been found to have metallicities in a range of $Z/Z_\odot=$ 1/10--1 
\citep[e.g.,][]{2007ApJ...661..115G,2009ApJ...695..580B,2010ApJ...715..656W,
2012ApJ...750..122B}. For these reasons, we attempted SED fitting with lower metallicities. 

The stellar opacity decreases with decreasing stellar metallicity, and thereby 
low metallicity increases the ionizing flux. Therefore, it may be inappropriate 
to adopt sub-solar metallicity, as it will lead to even lower 
SFR$_\mathrm{H\alpha}$/SFR$_\mathrm{FUV}$. 
Despite this concern, we performed independent additional SED fittings with 
fixed metallicities of $Z=0.004$, 0.008, and 0.02. 

As a result, the attenuation, especially in central regions, appeared to slightly increase 
with decreasing metallicity. However, as discussed earlier, a slight 
deviation in attenuation barely affects 
SFR$_\mathrm{H\alpha}$/SFR$_\mathrm{FUV}$. 

To quantitatively compare the results, we carried out the Kolmogorov--Smirnov 
test for the distributions of SFR$_\mathrm{H\alpha}$ and SFR$_\mathrm{FUV}$ 
inferred from different metallicities. It suggested that the null hypothesis that they 
are drawn from the same population cannot be rejected ($p$-value $\gtrsim0.98$). 

\subsection{LyC escape} \label{sec:disc:fesc}

Young, star-forming galaxies emit LyC radiation ($\lambda<912$\AA), and
the escape of these ionizing photons is considered to be one of the main 
contributors to the re-ionization of the early universe \citep[see][]{2009ApJ...706.1136O,
2009ApJ...693..984W,2011MNRAS.412..411Y,2015ApJ...811..140B,
2015MNRAS.451.2544P}. However, it is still debated how much of these photons 
escape from their host galaxies. Most studies have suggested that the 
LyC escape fraction of galaxies converges to below 0.15 \citep[e.g.,][]{2013A&A...553A.106L,
2014Sci...346..216B,2016MNRAS.461.3683I,2016ApJ...823...64L,
2019MNRAS.483.5223B}. In contrast, a few cases have been reported 
with high LyC escape fractions, up to 0.73, that might have originated from highly 
star-forming dwarf galaxies \citep[e.g.,][]{2017ApJ...837L..12B,
2018MNRAS.478.4851I,2019ApJ...878...87F}. 

The aforementioned studies were dealing with the entire light of individual 
galaxies, unlike this study focusing on local star-forming clumps. The ionizing 
photons are produced by very young stars surrounded by complex 
ISM structures \citep[e.g.,][]{1992ApJ...393..611W,2000ApJ...531..846D}. Stellar 
feedback can affect the dynamical state and morphology of the ISM, thereby 
instantaneously increasing the LyC escape fraction 
\citep[e.g.,][]{2018A&A...611A..95W,2020ApJ...902L..39B,2020ApJ...902...54C}. 

\begin{figure}[t]
\centering
\includegraphics[width=85mm]{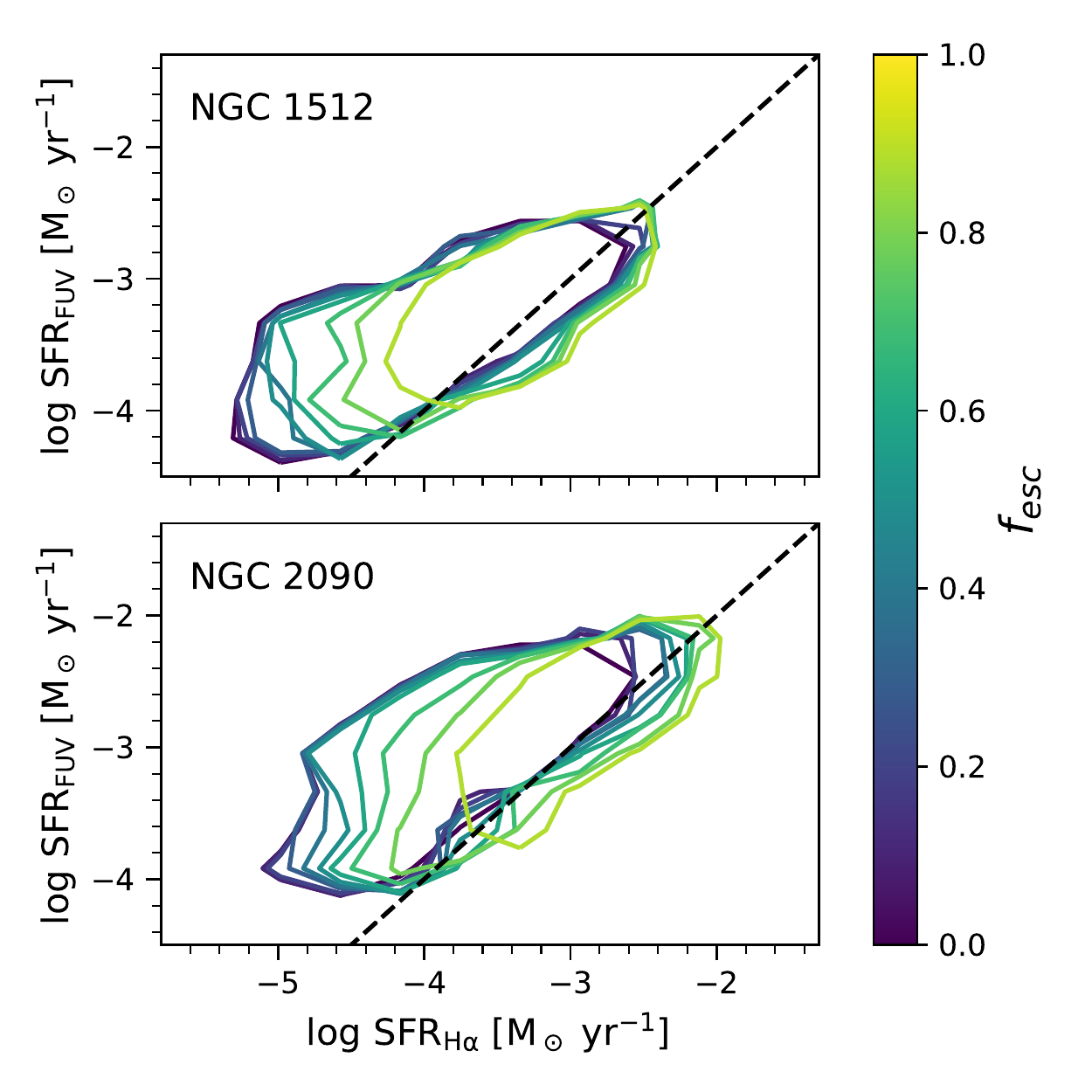}
\caption{Comparisons between the SFR$_\mathrm{FUV}$ and 
SFR$_\mathrm{H\alpha}$ of star-forming regions in NGC 1512 (top) and NGC 
2090 (bottom) with various upper limits of LyC escape fraction. Each contour 
contains data of more than 90\%, and is color-coded 
by LyC escape fraction. The dashed lines indicate a one-to-one relation 
between the SFRs. \label{fig:fig10}}
\end{figure}

Since the leakage of ionizing photons can result in an underestimation of 
SFR$_\mathrm{H\alpha}$, the observed H$\alpha$ flux must be corrected by 
the amount of loss using $f_\mathrm{H\alpha}^\mathrm{int} = 
f_\mathrm{H\alpha}^\mathrm{obs}/(1-f_{esc})$, where 
$f_\mathrm{H\alpha}^\mathrm{int}$ and $f_\mathrm{H\alpha}^\mathrm{obs}$ 
are the intrinsic and observed H$\alpha$ fluxes, respectively. 

The main results in this study were obtained using a LyC escape fraction of 
$f_{esc}=0.0$--0.3, but it is worthwhile to check whether the higher fractions can 
eliminate the discrepancy between the SFRs. We carried out additional 
SED fittings by allowing the upper limits of LyC escape fraction to vary up to 0.9 
while fixing the remaining parameters unaltered. 

\begin{figure*}[t]
\centering
\includegraphics[height=110mm]{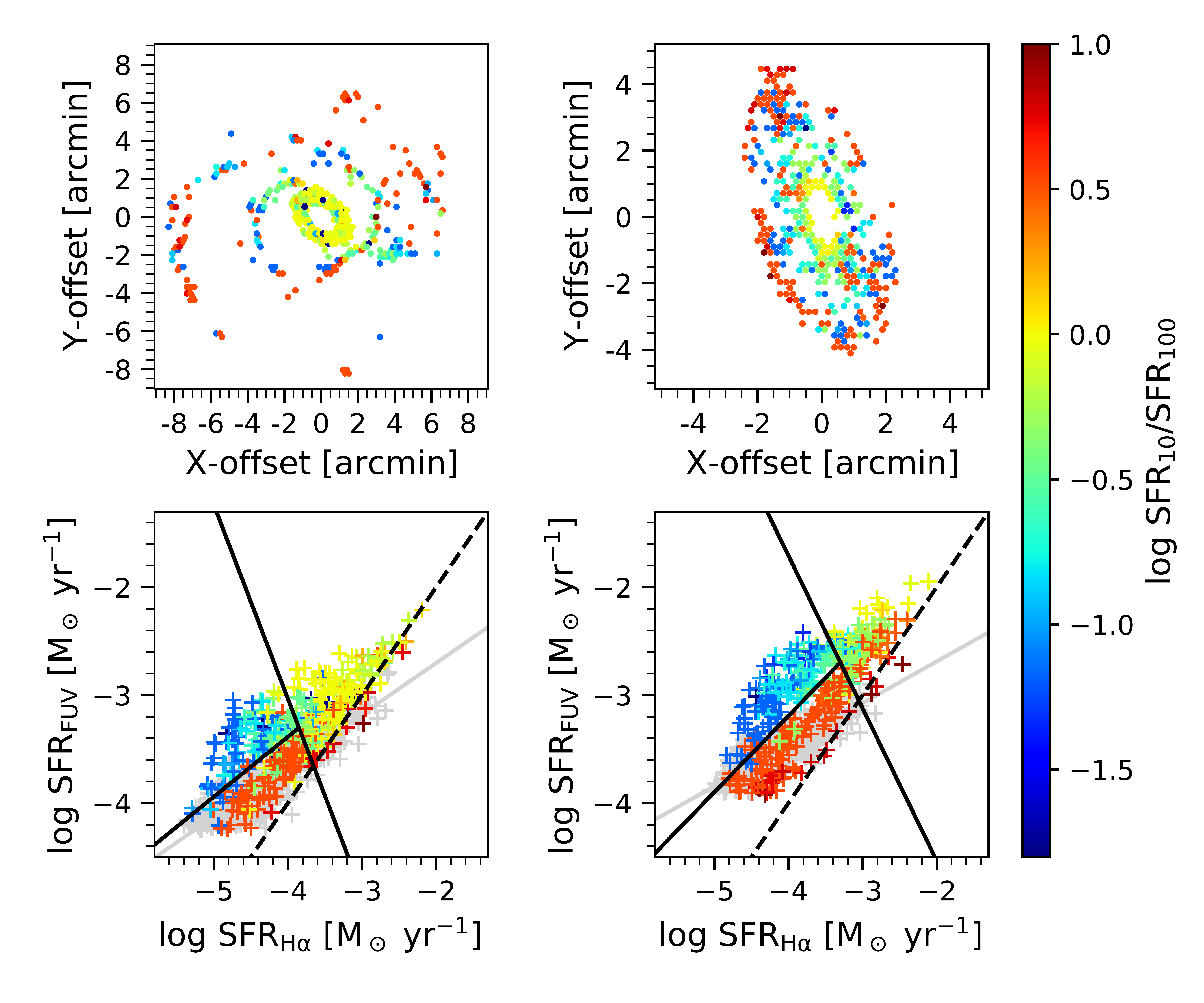}
\caption{Distributions of the ratio of average SFRs over the last 10 Myr and 100 Myr 
in star-forming regions of NGC 1512 (left) and NGC 2090 (right). It is presented in 
2-D maps (top) and SFR$_\mathrm{FUV}$ vs. SFR$_\mathrm{H\alpha}$ 
planes (bottom), respectively. The ratio of average SFRs over the last 10 Myr 
and 100 Myr is color-coded in the 
same range. The regions which appear to have different SFHs are divided by the 
black solid lines. The dashed lines 
indicate a one-to-one relation between the SFRs. The data uncorrected for the 
internal attenuation and LyC escape are shown in gray color.
\label{fig:fig11}}
\end{figure*}

Figure \ref{fig:fig10} shows the distributions of derived SFR$_\mathrm{FUV}$ 
and SFR$_\mathrm{H\alpha}$ when the loss of H$\alpha$ flux is fully recovered. 
For a better comparison, we applied the contours of different colors with a threshold 
containing $\gtrsim$90\% data. It is obvious that the contours approach a 
one-to-one relation between the SFRs as the LyC escape fraction increases. 
In addition, the contours gradually move to the top right direction as the 
attenuation from the newly-performed SED fitting is yielded to be slightly higher. 
When the LyC escape fraction is higher than 0.8, the 
discrepancy between the SFRs is marginally eliminated. However, it seems unlikely 
that most of the star-forming clumps have such an extremely high LyC 
escape fraction. This suggests that LyC escape alone cannot explain the low 
SFR$_\mathrm{H\alpha}$/SFR$_\mathrm{FUV}$ in the extended disks of galaxies. 

We also note that LyC photons can also be lost via dust absorption in ionized 
regions. However, the dust-to-gas mass ratio in ionized gas is likely to be 
substantially less than in neutral gas 
\citep[by a factor of 2 to 10;][]{2017MNRAS.469..630A}. It has also been found 
that dust has little effect on the nebular spectrum \citep{1986PASP...98..995M}. 
Scattering of H$\alpha$ can also lower the H$\alpha$ fluxes measured at 
star-forming regions, giving rise to the same effect as the LyC escape. However, 
the effect by dust scattering is, in general, more prominent at the FUV wavelengths, 
reducing the FUV flux more strongly. Most, if not all, of the single scattered light, 
will be lost from the aperture. On the other hand, some fraction of multiply 
scattered light can enter the aperture that is under consideration. However, 
the fraction of multiple-scattered light is reduced by a factor of $a^n$, where 
$a$ and $n$ are the dust albedo ($a < 1$) and number of scatterings, respectively. 
Thus, the net effect of dust scattering is the loss of light, unless the dust optical 
depth is so high that the observed flux is dominated by scattered light. As a 
result, it can further lower the H$\alpha$-to-FUV flux ratio when the amounts of loss in 
the H$\alpha$ and FUV fluxes are corrected, making the issue of the 
H$\alpha$ deficiency worse. Therefore, the dust absorption/scattering effects are not 
likely to significantly alter our results. 

\subsection{Star formation history} \label{sec:disc:sfh}

Since the H$\alpha$ and FUV fluxes trace star formation on different timescales, 
consistency between the estimated SFR$_\mathrm{H\alpha}$ and 
SFR$_\mathrm{FUV}$ can only be achieved when a constant SFR persists for at 
least 100 Myr. In other words, SFR variation in a shorter timescale may induce 
a discrepancy between SFR$_\mathrm{H\alpha}$ and SFR$_\mathrm{FUV}$. 

\begin{figure}[t]
\centering
\includegraphics[width=85mm]{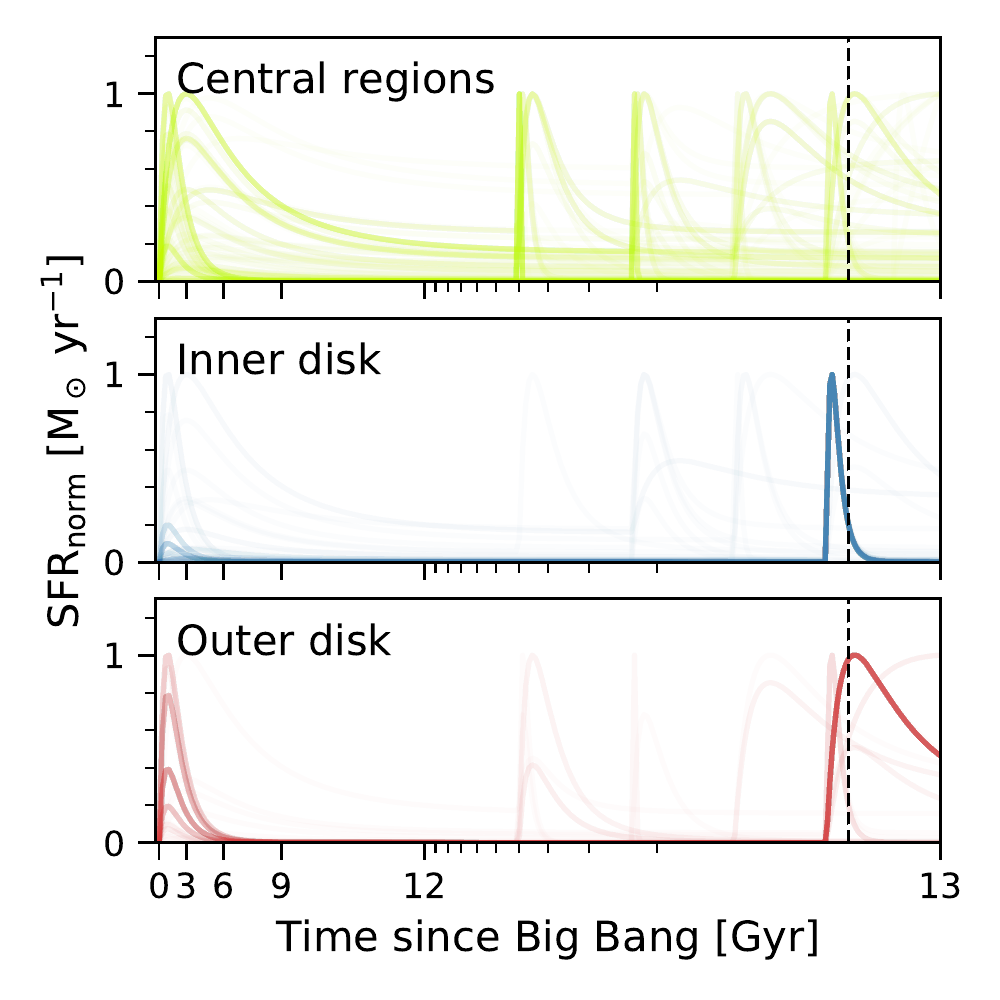}
\caption{SFHs for the star-forming regions in NGC 1512. The three groups as 
determined by SFR$_{10}$/SFR$_{100}$ in Figure \ref{fig:fig11} are presented 
separately. The curves denote SFHs for individual apertures, and the most 
vivid curves show the representative trends. The dashed lines represent a lookback 
time of 10 Myr. \label{fig:fig12}}
\end{figure}

\cite{2011ApJ...731...28A} investigated the stellar populations of five nearby 
XUV-disk galaxies using 
UV and NIR images. They found that the UV-bright sources in the outer 
disks of galaxies tended to have relatively old ages with a median of $\sim$100 Myr. 
Such a relatively old stellar population can lead to 
deficient H$\alpha$ emission in the outer disks. They also mentioned that this finding 
may be affected by biased sample selection since 
more than half of the sources had no H$\alpha$ emission. 
In contrast, we selected star-forming regions using both FUV and H$\alpha$, 
making our sample less biased compared to that of \cite{2011ApJ...731...28A}. 
Nevertheless, a deficit in H$\alpha$ flux in the extended disks of galaxies still appeared. 

A similar correlation between the H$\alpha$-to-FUV ratio and the H$\alpha$ (or FUV) 
intensity has been found in the diffuse regions outside of star-forming regions in 
external galaxies and our Galaxy \citep[e.g.,][]{2000ApJ...541..597H,2001ApJ...559..878H,
2011ApJ...743..188S}. They interpreted the trend using a scenario wherein the 
diffuse H$\alpha$ photons originate from relatively late-type stars (late O- and B-stars). 
Their interpretation (for the diffuse regions) is consistent with the present scenario 
(for the star-forming clumps) since the lifetime of early-type stars is shorter than 
that of late-type stars. 

Recently, \cite{2019ApJ...881...71E} described that starbursts with rising/falling 
phases can introduce a decline in the H$\alpha$/FUV luminosity ratio, and that 
this tendency is more distinct when the burst has a large amplitude and short 
duration. Indeed, the outer regions of XUV-disk galaxies are likely to have 
intermittent star-forming events because they have less gas and are vulnerable 
to stellar feedback. Hence, it would be useful to investigate when and where 
recent starbursts have occurred in understanding the origin of the deficient 
H$\alpha$ flux. 

To determine the phase of SFHs of late bursts, we adopted the CIGALE 
parameters of SFR$_{10}$ and SFR$_{100}$. This represents the average 
SFRs over the last 10 Myr and 100 Myr, respectively. We note that these are 
only tracers to check the SFHs, including recent starbursts, not proxies of 
SFR$_\mathrm{H\alpha}$ or SFR$_\mathrm{FUV}$. For a given e-folding 
time of the burst, we can expect SFR$_{10}$/SFR$_{100}$ to be lower than 
1 if the burst had started at a lookback time between 10 Myr and 100 Myr. 
Conversely, if the burst started within the last 10 Myr, the ratio would 
be higher than 1. Indeed, we allowed the age of the late burst to have sparse 
intervals (5, 15 Myr, and the rest) in the SED fitting to robustly distinguish the burst 
phase. Note that we attempted extra fittings with denser intervals. 
We found that even the detailed characteristics of late bursts vary, the overall trend 
almost matches the following descriptions. 

The upper panels in Figure \ref{fig:fig11} show the 2-D distributions of 
SFR$_{10}$/SFR$_{100}$ for the star-forming regions in the galaxies. 
Although SFR$_{10}$/SFR$_{100}$ are slightly mixed, 
it can be roughly divided into three groups according to the radial 
distance: the central regions, inner disk, and outer disk. The central regions 
seem to have SFR$_{10}$/SFR$_{100}\sim1$, indicating no or weak late bursts. 
The ratios of the inner disk appear to be lower than 1, while the outer disk has 
higher values. This implies that the burst populations in the inner disk tend to be 
older than those in the outer disk. Alternatively, the e-folding time of the inner disk 
might be shorter than that of the outer disk if the bursts had started simultaneously. 

In the bottom panels of Figure \ref{fig:fig11}, we present the distributions 
of SFR$_\mathrm{FUV}$ and SFR$_\mathrm{H\alpha}$, which are 
color-coded by SFR$_{10}$/SFR$_{100}$. The three groups defined above 
are clearly distinguishable on the 2-D planes. Note that not all data, roughly separated by 
three different color ranges, actually belong to three individual groups, but most of them do. 
We highlight that the discrepancy between the SFRs is maximized in the inner 
disk, where SFR$_{10}$/SFR$_{100}$ is the lowest. 

To investigate the overall SFHs for the three groups, we adopted a T-shaped 
boundary consisting of the linear fit and its orthogonal function. Each intersection 
was determined arbitrarily by visual inspection. Figure \ref{fig:fig12} shows the 
SFHs reproduced with the CIGALE results for three groups in NGC 1512. 
In order to emphasize the representative trend in each group, we normalized 
the SFRs by the highest value in each SFH, and applied translucent colors. 
As we mentioned above, the central regions were dominated by the main populations, 
which can be described as an exponential SFH with insignificant late bursts 
over the last 100 Myr. In contrast, 
the inner and outer disks clearly appear to have had late bursts. Interestingly, 
almost all of these late bursts started at the same time ($t_\mathrm{lookback}=15$ Myr), 
but they have evolved with different e-folding times. 

We also note that the maximum amplitudes of the main populations in the inner 
disk tend to be smaller than those of the outer disk. In summary, the late bursts 
in the disks occurred in two different types: (1) bursts with strong amplitude and 
short e-folding time in the inner disk, and (2) bursts with moderate amplitude and 
long e-folding time in the outer disk. Together with the descriptions of 
\cite{2019ApJ...881...71E}, we conclude that the strong discrepancy between 
SFR$_\mathrm{H\alpha}$ and SFR$_\mathrm{FUV}$ in the inner disk could be 
induced by the rapidly decreasing SFR over the last 10 Myr. 

As we mentioned in Section \ref{sec:data:sample}, the two galaxies are thought 
to reside in different environments. This leads to the inference that the mechanisms 
triggering SF in the extended disks might also be different. For example, NGC 1512 
was expected to exhibit signs such as asymmetric and delayed SF. However, the 
SFHs, including late bursts in the extended disks of both galaxies, were revealed 
to show quite azimuth-symmetric distributions. In other words, the late bursts in 
the extended disks had started simultaneously, and then underwent an ``inside-out'' 
suppression. Consequently, these findings suggest that the internal processes of 
galaxies or gas accretion from the intergalactic medium might be responsible for 
the recent SF in the extended disks of galaxies. This interpretation is still 
inconclusive because of the small sample size. We expect that larger samples of 
XUV-disk galaxies would be helpful to understand the origin of recent SF in galaxy 
outskirts. 

\begin{figure}[t]
\centering
\includegraphics[width=85mm]{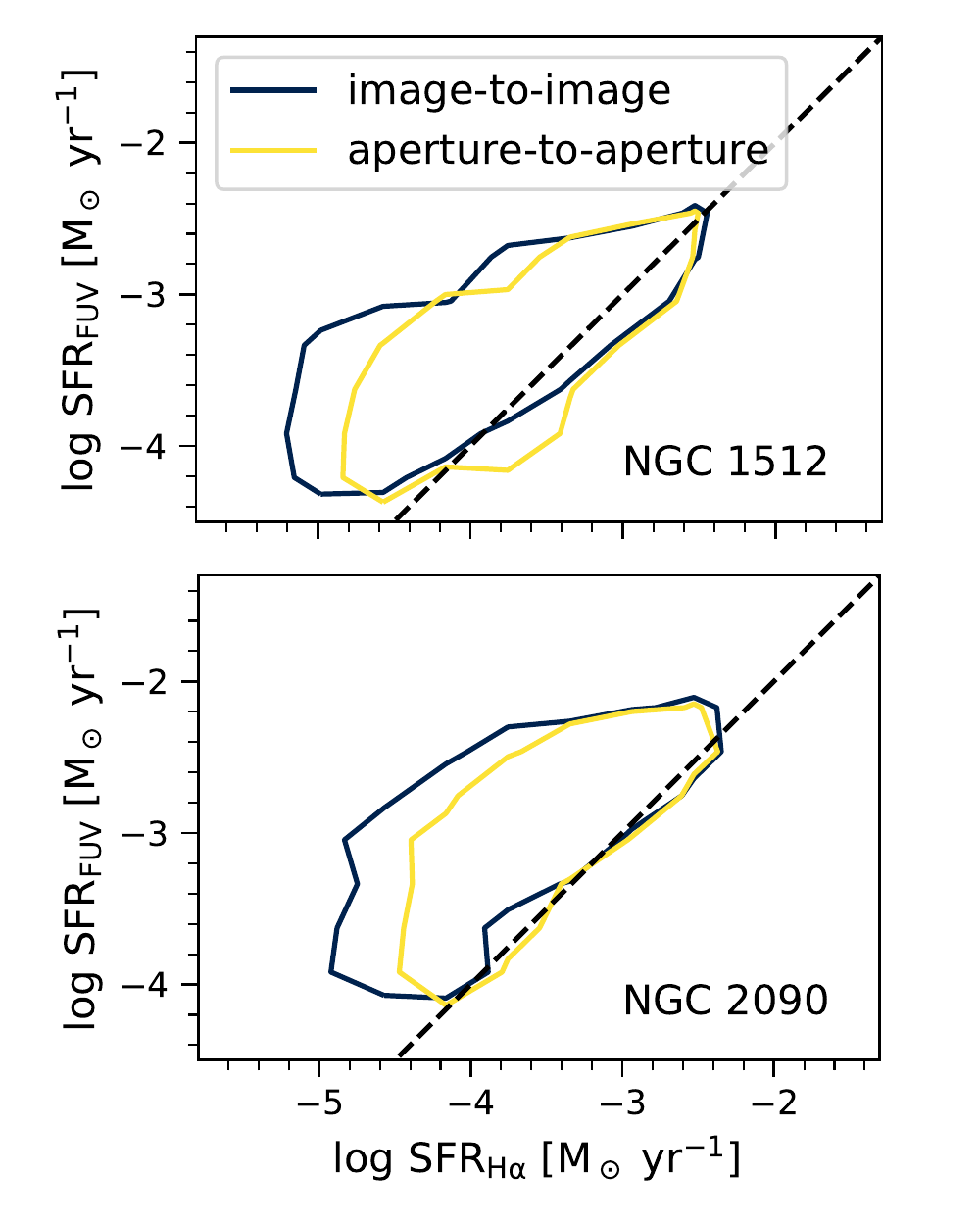}
\caption{Comparisons of the SFR$_\mathrm{FUV}$ and SFR$_\mathrm{H\alpha}$ 
of the star-forming regions in NGC 1512 (top) and NGC 2090 (bottom). 
It is color-coded to indicate which method was used to subtract the stellar 
continuum: `image-to-image' and `aperture-to-aperture'. 
Each contour contains data of more than 90\%. The dashed 
lines indicate a one-to-one relation between the SFRs. \label{fig:fig13}}
\end{figure}

\subsection{Uncertainty of the H$\alpha$ flux calibration} \label{sec:disc:BNCR}

All of the above results are based on H$\alpha$ fluxes measured from 
continuum-subtracted H$\alpha$ images. These images were created by 
`image-to-image' subtraction of the $R$-band images scaled by a single BNCR 
from the H$\alpha$ narrowband images. In Section \ref{sec:data:redux}, we 
demonstrated that BNCR depends on the $g-r$ color of objects, and the individual 
star-forming clumps in galaxies have various colors. For this reason, we may 
concern that using a single BNCR would cause an inaccurate continuum subtraction. 
That is, the choice of a minimum BNCR can give rise to an 
over-subtraction of the continuum in the regions with bluer colors. 

Alternatively, the H$\alpha$ fluxes can be estimated using `aperture-to-aperture' 
subtraction for the H$\alpha$ narrowband and $R$-band fluxes measured in 
individual apertures. In this approach, each $R$-band flux is scaled by the 
corresponding BNCR for each aperture. In order to examine whether this method 
resolves the discrepancy between the SFRs, we attempted to re-analyze using the 
H$\alpha$ fluxes measured in this way. 

Figure \ref{fig:fig13} compares the results of `image-to-image' and `aperture-to-aperture' 
continuum subtraction. Here, the contours represent a threshold containing 
$\gtrsim$90\% data. The results for low SFR regions changed significantly depending 
on the adopted subtraction method, while it was not significantly altered in high SFR 
regions. This is a natural consequence as the regions with high SFRs, especially the 
central regions, tend to have redder colors. However, the discrepancy between the 
SFRs still exists even if `aperture-to-aperture' subtraction is an appropriate method for 
estimating the H$\alpha$ flux. This suggests that the lack of H$\alpha$ flux in the extended 
disks of galaxies is an intrinsic phenomenon, rather than introduced by incorrect flux 
measurement. 

\section{Summary} \label{sec:summary}

Using the H$\alpha$ images obtained with KMTNet, we examined the characteristics 
of SF for two nearby spiral galaxies with extended UV features: NGC 1512 and NGC 
2090. The main goal of this study was to study the discrepancy between the SFRs 
traced by H$\alpha$ and FUV fluxes of star-forming regions in galaxies. In addition, 
we inferred the SFHs in the extended disks of galaxies, taking advantage of the fact 
that the H$\alpha$ and FUV trace SFRs of different timescales. 

For a spatially resolved analysis, we conducted grid-shaped aperture photometry 
using an aperture size comparable to the size of typical H\textsc{ii} regions. 
Through an SED fitting with CIGALE, we examined several important factors 
that can affect the estimation of H$\alpha$ and FUV fluxes. Finally, we compared 
SFR$_\mathrm{H\alpha}$ and SFR$_\mathrm{FUV}$, which were estimated from 
the intrinsic H$\alpha$ and FUV fluxes, respectively. The two SFRs appeared to be 
different, and the discrepancy between them became severe in the low SFR regime. 

We evaluated the possible causes of this discrepancy. Firstly, we assessed the 
effect of uncertain internal attenuation and metallicity. Both parameters can affect 
estimates of H$\alpha$ and FUV fluxes, but their effects on the resultant SFRs 
were negligible. We also examined how much ionizing photons should escape 
from individual star-forming clumps to eliminate the discrepancy between the SFRs. 
This revealed that the LyC escape fraction needs to be higher than 80\% in the 
extended disks of galaxies. Therefore, we concluded that LyC escape would be 
only a secondary factor at best because it is unlikely that almost all of the 
star-forming clumps would have such high LyC escape fractions. We also traced 
recent bursts in individual star-forming clumps using SFR$_{10}$/SFR$_{100}$. 
This revealed that the regions exhibiting a significant discrepancy between the 
SFRs tend to have SFHs with intense starbursts, which have been in a 
rapidly-falling phase in the last 10 Myr. Therefore, we suggest that recent 
starbursts with strong amplitude and short e-folding time are likely a primary 
factor for the deficit of H$\alpha$ flux. 

In addition, we inspected whether continuum subtraction was appropriately 
performed. The results indicated that our analyses might have been affected by an 
underestimation of H$\alpha$ flux. However, the discrepancy between the SFRs 
still existed regardless of this issue. 

From the spatially resolved analysis on SFHs, we found that the recent starbursts 
in the extended disks of galaxies had occurred simultaneously and 
azimuth-symmetrically. The causes of the recent SF doesn't seem to be closely 
related to the presence or absence of interacting companions. This suggests that 
the recent starbursts may have been triggered by gas accretion or internal origins 
rather than external perturbation. As the KMTNet Nearby Galaxy Survey 
continues, we expect that larger samples of XUV-disk galaxies will 
be helpful to understand the origin of recent SF in the extended disks. 

\acknowledgments
We are grateful to an anonymous referee for informative comments and suggestions. 
We thank Taysun Kimm for the fruitful discussions. This research has made use of the 
KMTNet system operated by the Korea Astronomy and Space Science Institute (KASI) 
and the data were obtained at CTIO in Chile, SAAO in South Africa, and SSO in Australia. 
YKS acknowledges support from the National Research Foundation of Korea (NRF) grant 
funded by the Ministry of Science and ICT (NRF-2019R1C1C1010279). LCH was supported 
by the National Science Foundation of China (11721303, 11991052, 12011540375) and the 
National Key R\&D Program of China (2016YFA0400702). This work was supported by the 
National Research Foundation of Korea (NRF) grant funded by the Korea government (MSIT) 
(No.2020R1A2C4001753) and under the framework of international cooperation program 
managed by the National Research Foundation of Korea (NRF-2020K2A9A2A06026245).

%

\vspace{5mm}
\facilities{KMTNet, GALEX, Spitzer, AAVSO, IRSA, NED}


\software{Astropy \citep{astropy:2013,astropy:2018}, Scipy \citep{2020SciPy-NMeth}, 
SExtractor \citep{1996A&AS..117..393B}, Photutils \citep{larry_bradley_2021_4453725},
CIGALE \citep{2019A&A...622A.103B}, SCAMP \citep{2006ASPC..351..112B}, 
SWarp \citep{2002ASPC..281..228B}, Matplotlib \citep{Hunter:2007}
}




\bibliography{ms}{}
\bibliographystyle{aasjournal}



\end{document}